\documentclass[twocolumn, dvipsnames]{aastex701}

\usepackage{amsmath}
\usepackage{amssymb} 
\usepackage{xspace}
\usepackage{color}
\usepackage{comment}
\usepackage{csquotes}
\usepackage{placeins}
\usepackage{gensymb}
\usepackage{soul}

\defcitealias{Carrasco2010}{C10}
\defcitealias{Carrasco_dissecting_2021}{C21}
\defcitealias{zenteno20}{Z20}
\defcitealias{Zenteno_BCG_2025}{Z25}
\defcitealias{Kluge_optical_2024}{K24}

\shorttitle{Merging Galaxy Cluster Candidate}
\shortauthors{Aradhey et al.}

\begin{document} 

\title{The extreme Disturbed Galaxy Cluster Sample (eDGeS): Spectroscopy and Dynamical Analysis of the eROSITA-Selected Merging Galaxy Cluster Candidate 1eRASS J025859.4-193727}

\author[orcid=0009-0003-7308-9413]{Anish S. Aradhey}
\affiliation{Department of Physics and Astronomy, University of North Carolina at Chapel Hill, 120 E. Cameron Ave., Phillips Hall CB3255, Chapel Hill, NC 27599, USA}
\email[show]{aaradhey@unc.edu}

\author[orcid=0000-0002-7272-9234]{Eleazar R. Carrasco}
\affiliation{International Gemini Observatory, NSF 
NOIRLab, Casilla 603, La Serena, Chile}
\email[show]{rodrigo.carrasco@noirlab.edu}

\author[orcid=0000-0001-6455-9135]{Alfredo Zenteno}
\affiliation{Cerro Tololo Inter-American Observatory, NSF 
NOIRLab, Casilla 603, La Serena, Chile}
\email[]{alfredo.zenteno@noirlab.edu}

\author[orcid=0000-0001-6419-8827]{Rog\'erio Monteiro-Oliveira}
\affiliation{Observat\'orio Nacional (ON/MCTI), Rua General Jos\'e Cristino, 77, Rio de Janeiro 20921-400, Brazil}
\email[]{rmo@on.br}

\author[orcid=0000-0003-4232-8584]{Facundo A. G\'omez}
\affiliation{Departamento de Astronom\'ia, Universidad de La Serena, Av. Ra\'ul Bitr\'an 1305, La Serena, Chile}
\email[]{fagomez@userena.cl}

\author[orcid=0000-0003-4432-5037]{Konstantina Boutsia}
\affiliation{Cerro Tololo Inter-American Observatory, NSF
NOIRLab, Casilla 603, La Serena, Chile}
\email[]{konstantina.boutsia@noirlab.edu}

\author[orcid=0000-0002-7619-5399]{Esra Bulbul}
\affiliation{Max Planck Institute for Extraterrestrial Physics, Giessenbachstrasse 1, 85748 Garching, Germany}
\email[]{ebulbul@mpe.mpg.de}

\author[orcid=0000-0002-2651-7038]{Guillermo Damke}
\affiliation{Cerro Tololo Inter-American Observatory, NSF NOIRLab, Casilla 603, La Serena, Chile}
\email[]{guillermo.damke@noirlab.edu}

\author[orcid=0000-0003-0506-214X]{Amelia Ramirez}
\affiliation{Departamento de Astronom\'ia, Universidad de La Serena, Av. Ra\'ul Bitr\'an 1305, La Serena, Chile}
\email[]{aramirez@userena.cl}

\author[orcid=0000-0002-1397-4045]{Hector Cuevas}
\affiliation{Departamento de Astronom\'ia, Universidad de La Serena, Av. Ra\'ul Bitr\'an 1305, La Serena, Chile}
\email[]{hcuevas@userena.cl}

\begin{abstract}
Addressing open questions such as the effect of galaxy cluster mergers on the evolution of galaxies and the self-interaction cross-section of dark matter requires the construction of a large, diverse sample of disturbed galaxy clusters. One contribution to such a sample is the extreme Disturbed Galaxy Cluster Sample (eDGeS), an ongoing spectroscopic study of massive disturbed and merging galaxy clusters selected from complementary surveys. Here, we present our first results from the eROSITA-selected subsample (eDGeS-eROSITA): a dynamical state analysis for the merging cluster candidate 1eRASS J025859.4-193727.
We present the first spectroscopic redshift measurements for 76 galaxies in the direction of the cluster, confirming 51 as cluster members. The cluster displays a line-of-sight velocity dispersion of $\sigma_{\rm{cl}}=808^{+59}_{-66}~\rm{km~s^{-1}}$ and a dynamical mass of $M_{200}=4.47^{+0.94}_{-0.93} \times 10^{14}~M_{\odot}$. While the cluster shows no evidence for substructure along the line of sight, the projected distribution of member galaxies reveals three subclusters, only one of which appears associated with the cluster’s X--ray contours. Phase-space analysis reveals a large fraction of recently infallen member galaxies. Together, this evidence suggests a cluster merger along the plane of the sky. Assuming that the two most massive subclusters are on outgoing (incoming) trajectories after their first pericentric passage, a two-body Monte Carlo model predicts that the two subclusters collided $0.50^{+0.20}_{-0.22}~\rm{Gyr}$ ($1.24^{+0.96}_{-0.38}~\rm{Gyr}$) ago.
\end{abstract}


\section{Introduction}\label{sec:intro}



Galaxy clusters are the most massive virialized systems in the universe.
These structures occupy the densest peaks of the dark matter (DM) field and grow through a combination of smooth accretion along filaments and mergers of smaller groups and clusters \citep{nelson24}, processes that are intrinsically stochastic and highly energetic \citep{Sarazin04}. The constant accretion implies that these structures are rarely in equilibrium and experience major and minor mergers during their assembly history \citep{KravtsovBorgani2012}. These systems can exist in various dynamical states, ranging from systems close to virialization \citep[e.g.][]{Carrasco2010, Soja18, Watson2026} to highly perturbed systems that are still undergoing formation \citep[e.g.][]{Monteiro-Oliveira_A1758_2017, Monteiro-Oliveira17b, Monteiro-Oliveira_A2034_2018, Monteiro-Oliveira_A1644_2020, Monteiro-Oliveira22-eROSITA, Lourenco_bullet_2020}.

Galaxy clusters in different evolutionary stages offer unique and fundamental insights into galaxy evolution and cosmological structure formation.
For example, dynamically relaxed galaxy clusters, characterized by regular X--ray morphologies and intracluster media (ICM) close to hydrostatic equilibrium, play a central role in cluster cosmology, providing the primary calibration for X--ray mass proxies and establishing scaling relations used in cosmological studies \citep[e.g.][]{Kravtsov2006, Vikhlinin2009, Pratt2019, Monteiro-Oliveira21, Doubrawa2023, Okabe_WL_2025, Ding_miscenter_2025, Chiu_WLcalib_2025, Aldas_scaling_2026}. Moreover, since no obvious substructures are detected in the optical or X--ray, relaxed galaxy clusters are ideal systems for studying dark matter density profiles and mass distributions by combining gravitational lensing (weak and/or strong) with dynamical measurements \citep[e.g.,][]{Monna2017,Verdugo2020,Carrasco_dissecting_2021,Monteiro-Oliveira21}.

Merging galaxy clusters provide unique laboratories for several purposes. For example, observations of merging clusters and their relative velocities may serve as powerful constraints for cosmology and dark energy models \citep[see, e.g.,][]{Bouillot_pairwise_2015}. Additionally, mergers provide information on the physical properties of the ICM \citep[see][for a review]{Markevitch07} because they dissipate large amounts of kinetic energy into the ICM, generating shocks, turbulence, and complex thermodynamic structures \citep[e.g.,][]{Owers2014, Molnar16}.


Merging galaxy clusters are also important for probing the nature of dark matter \citep[see][for a recent review]{Monteiro-Oliveria_Review_2026}, as cluster mergers can produce spatial offsets between the collisional (gas) and effectively collisionless (galaxies and dark matter) cluster components. Such offsets provide direct constraints on the dark matter self-interaction cross section \citep[e.g.,][]{Markevitch2004, Randall2008, Harvey2013, Harvey2015, Wittman_mismeasure_2018, Fischer2023}. 
In particular, collisions with low impact parameters and occurring primarily in the plane of the sky maximize the observable separation between components and minimize projection effects, making them especially sensitive probes of self-interacting dark matter \citep[SIDM, e.g.][]{ Randall2008, Tam_SIDM_2026}.

Finally, merging galaxy clusters offer a unique window into galaxy evolution in extreme environments \citep[e.g.,][]{zenteno20, veliz2025, Ahad26}. During cluster mergers, galaxies are subject to strong ram-pressure stripping, producing dramatic features such as jellyfish galaxies \citep[e.g.,][]{McPartland2016, Roberts2022, Mello-Terencio_wind_2026}. While these provide clear evidence of environmental processing, the global impact of cluster mergers on galaxy populations remains less well established \citep[see, e.g.,][]{aldas2023}. Some studies report enhanced star formation activity triggered by merger-driven processes \citep[e.g.,][]{hernandez-lang22, aldas2023, aldas2025}, whereas others find little or no significant difference relative to galaxies in relaxed clusters \citep[e.g.,][]{Mansheim17}. More recent analyses suggest that both behaviors may coexist, depending on the merger stage and the ability to disentangle merger-driven effects from those associated with the preexisting cluster environment \citep[e.g.,][]{Kelkar2020, Lourenco2023, PirainoCerda2024}. 


In order to generalize these findings, the aforementioned analyses must be applied to a larger, statistically significant sample of merging systems. Thus,
it is essential to expand the census of well-characterized merging clusters, spanning different merger configurations (e.g., mass ratios, impact parameters) and stages (i.e., time since or until collision). Such samples are becoming increasingly larger \citep[e.g.,][]{wenhan13, zenteno20, Yuan22, Wen_catalog_2024, Sanders_morphologies_2025, Ahad26}.
To expand the aforementioned sample, we have established the extreme Disturbed Galaxy cluster Sample (eDGeS), a spectroscopic study designed to investigate the dynamical state and evolution of massive disturbed clusters selected from complementary surveys. Here, we present the first results from the eROSITA-selected subsample (eDGeS-eROSITA).

The eDGeS-eROSITA subsample draws on the results of \citet[][\citetalias{Zenteno_BCG_2025} hereafter]{Zenteno_BCG_2025}, who classified the dynamical state of 3,946 of the X--ray selected galaxy clusters from the first SRG/eROSITA Western All-Sky Survey \citep[eRASS1,][]{bulbul24} using the offset $D_{\rm BCG-X}$ between the Brightest Cluster Galaxy (BCG) identified optically by 
\citet[][\citetalias{Kluge_optical_2024} hereafter]{Kluge_optical_2024}
and the X--ray peak/centroid. The $D_{\rm BCG-X}$ is a metric adopted as a reliable indicator of the dynamical state of galaxy clusters \citep[e.g.,][]{mann12, lopes18, zenteno20} and is particularly efficient 
when relying on shallow optical and X--ray imaging.  

\begin{figure*}
    \centering
    \includegraphics[width=0.8\linewidth]{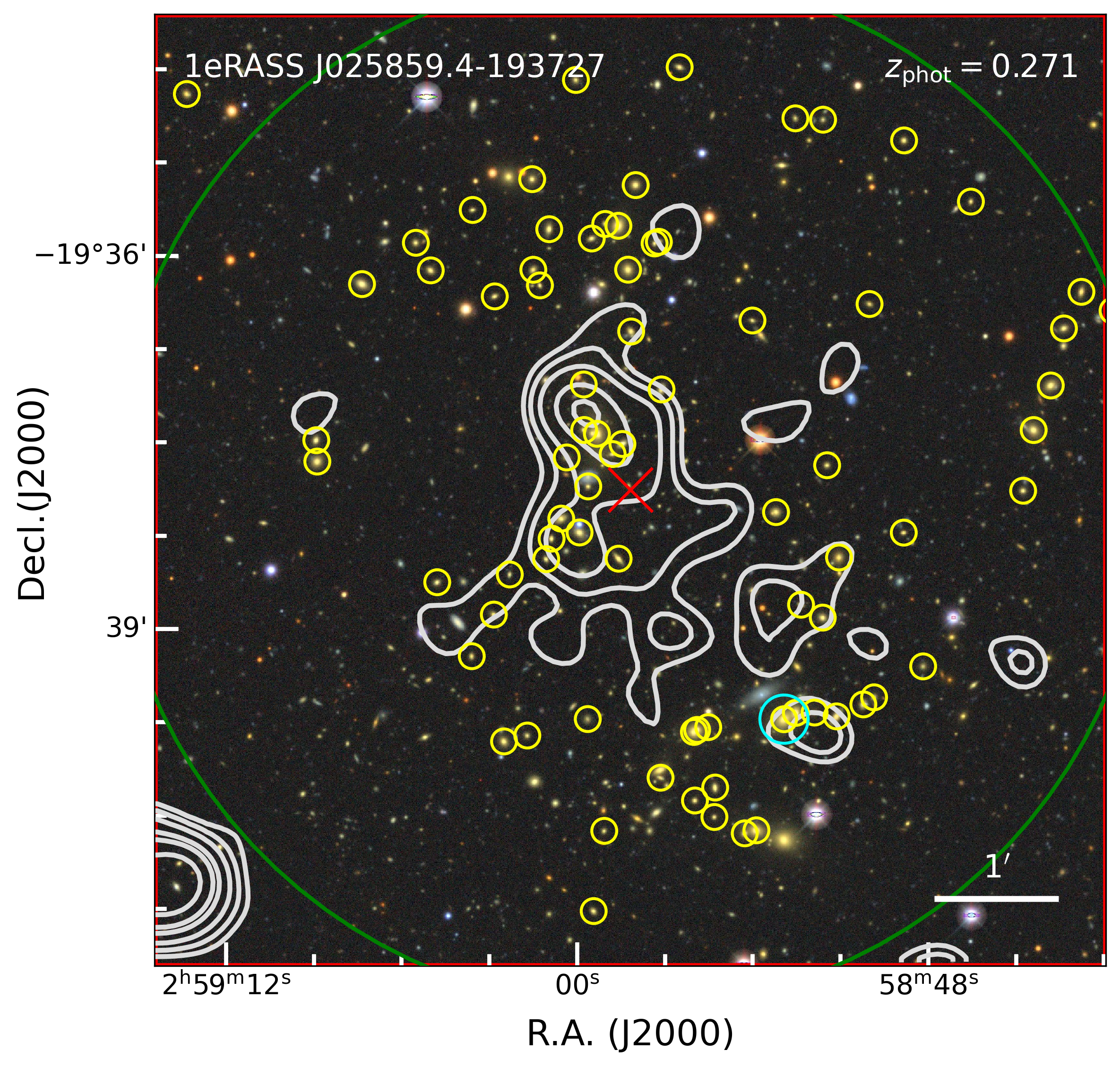}
 \caption{Legacy Survey color-composite (red-green-blue) image of the galaxy cluster J025859.4-193727. The gray contours correspond to eROSITA DR1 X--ray surface brightness contours (in counts s$^{-1}$ pixel$^{-1}$) smoothed using a Gaussian of 12\arcsec. The levels correspond to 80, 90, 95, 98, 99, and 99.5 percentiles. The small yellow circles denotes the photometrically-selected cluster members ($P_{\rm mem}>0.5$), 
 and the cyan circle marks the candidate cluster BCG, both from \citetalias{Kluge_optical_2024}. The green circle denotes the $R_{500}$ radius. The red \enquote{X} shows the position of the refined X--ray centroid from \citet{Bulbul_clusters_2024} (\texttt{RA\_XFIT, DEC\_XFIT}). The size of the image is $8.2 \times 8.2$ arcmin$^2$ ($\sim 2 \times 2$ Mpc$^2$). North is up, and east is to the left.}
 
    \label{fig:cluster_inspector}
\end{figure*}

Among the \citetalias{Zenteno_BCG_2025} sample, 
\object{1eRASS J025859.4-193727} (hereafter J025859 for short, Fig. \ref{fig:cluster_inspector}) stood out as a particularly promising merger candidate. First cataloged as Abell 406 \citep[][offset $\sim$3\arcmin\ from the eROSITA detection]{Abell_catalog_1989}, the cluster's morphology was noted early on for its elongation, wide distribution of brightest galaxies, and multiple concentrations of galaxies \citep{Abell_catalog_1989}, which together suggest multiple subcluster components at similar distances \citep{Struble_Morphological_1987, Struble_Binary_1958}. This unrelaxed optical morphology was confirmed by \citet{Wen_DESI_2024} (cataloged as WH J025859.3-193726, $\sim$2\arcsec\
offset from the eROSITA detection), who calculated the relaxation parameter $\Gamma <0$  based on the smoothed, luminosity-weighted projected distribution of member galaxies \citep{wenhan13}. Other modern photometric studies (\citealp{Wen_SuperCOSMOS_2018, Zou_DESI_2021, Wen_unWISE_2022, Wen_DESI_2024}; \citetalias{Kluge_optical_2024}; \citealp{Klein_ACTMCMF_2024}) have estimated the cluster's redshift ($\overline{z}\approx 0.242-0.287$), radius ($R_{500} \approx 1.041 - 1.194~\rm{Mpc}$), and mass ($M_{500} \approx 4.37-5.58 \times 10^{14}~\rm{M_{\odot}}$).

In addition to optical detections, the X--ray characterization of J025859 revealed the large $D_{\rm BCG-X}$ value \citepalias{Zenteno_BCG_2025} and disturbed X--ray morphology \citep{Sanders_morphologies_2025}. 
Additionally, X--ray observations yielded estimates of $R_{500} = 1.065^{+0.027}_{-0.031}~\rm{Mpc}$ and $M_{500} = 4.53^{+0.36}_{-0.38} \times 10^{14}~M_{\odot}$ \citep{Bulbul_clusters_2024}. Furthermore, the cluster has been detected via the SZ effect with the Atacama Cosmology Telescope as ACT-CL J0258.9-1938 \citep[][offset $\sim$1.4\arcmin\ from the eROSITA detection]{Hilton_ACT_2021}, resulting in estimates of $M_{500} = 3.96^{+1.03}_{-0.86} \times 10^{14}~M_{\odot}$ and $M_{200} = 5.22^{+1.26}_{-1.02} \times 10^{14}~M_{\odot}$. The aforementioned optical photometric estimates agree well with these X--ray and SZ-derived parameters.

Despite these investigations across multiple wavelengths, the cluster remains severely understudied spectroscopically, as the literature lacks any published redshifts for the galaxies in the immediate field of the cluster. Thus, our paper presents the first spectroscopic dynamical analysis along with the first catalog of spectroscopic redshifts for member galaxies.
The paper is organized as follows. Section~\ref{sec:data} provides details on 
spectroscopic observations, data reduction, and redshift estimation of the candidate cluster members. In Section~\ref{sec:analysis}, we 
calculate the cluster's dynamical parameters, 
including redshift and mass. In Section~\ref{sec:substructure}, we identify and characterize cluster substructures. In Section~\ref{sec:collision}, we utilize a Monte Carlo simulation to reconstruct the cluster's collision scenario.
We discuss our findings in Section~\ref{sec:discussion}. Finally, we summarize our results and present our conclusions in Section~\ref{sec:conclusion}.
Throughout this work, we adopt a flat $\Lambda$CDM cosmology with
$H_0$ = 68.3 km s$^{-1}$ and $\Omega_{M} = 0.299$ \citep{Bocquet_cosmology_2015}.
With this cosmology, 1\arcsec\ corresponds to 4.21 kpc at the redshift of the cluster ($\overline{z} \approx 0.267$). All magnitudes presented in this paper are quoted in the AB system.

\section{The Data}
\label{sec:data}

\subsection{Target Selection}
\label{subsec:selecttargets}

\begin{figure}
    \centering
    \includegraphics[width=1.0\linewidth]{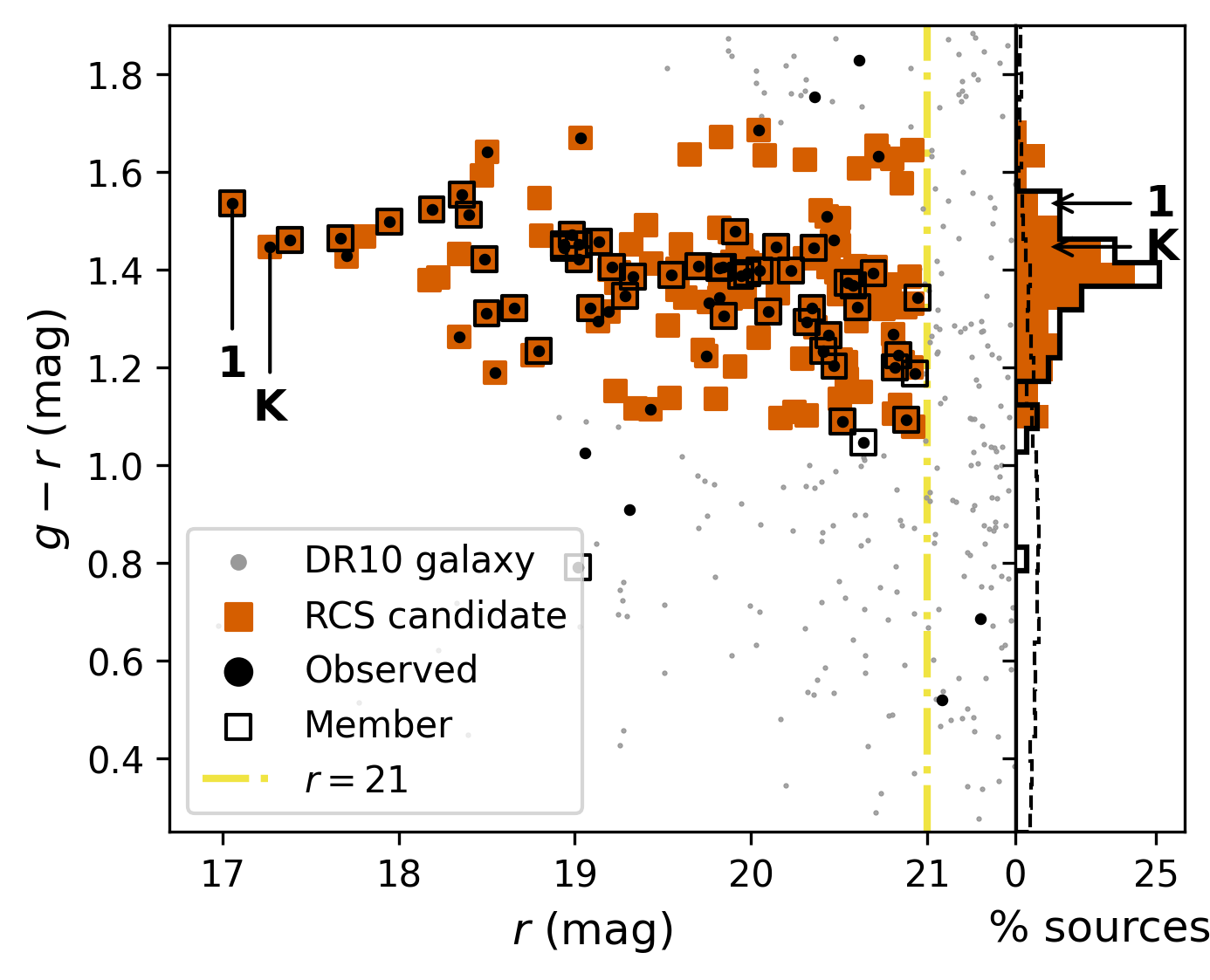}
    \caption{
     Color–magnitude diagram for all LSDR10 galaxies 
     within $\sim$1.5$R_{500}$ ($\sim$0.2$^{\circ}$). The filled red squares represent galaxies within the fitted red cluster sequence with $r\leqslant21$ (yellow dash-dotted line). 
     Black bullets represent the observed galaxies, while black square borders show the spectroscopically confirmed members.
     The brightest spectroscopically confirmed member and the BCG photometrically selected by \citetalias{Kluge_optical_2024} are labeled with \enquote{1} and \enquote{K,} respectively. The right panel shows the distribution of dereddened $g-r$ color for all LSDR10 galaxies (black dashed line), RCS candidates (red solid bars), and spectroscopically confirmed members (black solid line).
    }
    \label{fig:member_CMD}
\end{figure}

Here we present a brief overview of the candidate member galaxies selected for spectroscopic follow-up. To maximize our chances of observing galaxies that are cluster members, we selected targets for spectroscopic observations using the linear color-magnitude relation for early-type galaxies in clusters, i.e., the red cluster sequence \citep[RCS,][]{Gladders_sequence_1998, Gladders_method_2000}. 
To construct the RCS, we first selected all galaxies around the cluster center with
membership probability $P_{\rm mem}>0.5$ calculated by \citetalias{Kluge_optical_2024} 
using the RCS-based \texttt{eROMaPPer} algorithm\footnote{The catalog of galaxies with $P_{\rm mem}$ was retrieved from: https://erass-cluster-inspector.com \citepalias{Kluge_optical_2024}}.
We then constructed a $g-r$ vs. $r$ color-magnitude diagram (CMD) 
using the entries for these high-membership-probability galaxies in the Legacy Surveys Data Release 10 (LSDR10) Main Tractor Photometry catalog (\texttt{ls\_dr10.tractor}\footnote{datalab.noirlab.edu/data-explorer?showTable=ls\_dr10.tractor}) and defined the RCS as the line in the CMD that best fits these galaxies. The LSDR10 catalog was built including data from DR9 \citep{dey19}, DES \citep{DES21A}, DELVE \citep{drlica-wagner22}, DeROSITAS \citepalias{Zenteno_BCG_2025}, and several other smaller DECam programs. 
Next, we 
selected all LSDR10 galaxies with $r$-band magnitude $r\leqslant24$ within a radius of $1.5 R_{500}$ ($\sim$0.2$^\circ$) around the eROSITA cluster coordinates. All galaxies located within $\pm 2 \sigma$ from the best-fit RCS and brighter than the characteristic magnitude $m^{*}+2.0$ ($r \lesssim 21$) were selected as potential targets for spectroscopy. More details on the selection of the candidates will be presented in a future eDGeS paper. Overall, 77 galaxies were selected and included in two multi-object masks for spectroscopic observations.
The CMD of galaxies lying in the RCS and the observed galaxies are shown in Figure \ref{fig:member_CMD}.

\subsection{Optical Spectroscopic Data}
\label{subsec:spectra}
 
Spectroscopic data were obtained with the Gemini Multi-Object Spectrograph \citep[hereinafter GMOS;][]{Hook_GMOS_2004} mounted at the Gemini South telescope in Chile, in queue mode. GMOS observations were carried out on 2024 September 27 UT (Program ID: GS-2024B-Q-317, PI: Carrasco), during dark time and clear skies, under average seeing conditions ($\simeq$ 0.7\arcsec - 0.8\arcsec), and with an airmass between $\sim$1.09 and $1.25$.

All spectra were acquired using the B480 grating, the GG455 blocking filter, 1\arcsec\ slits, and 2$\times$2 binning during three 600-second exposures to obtain a final $S/N\gtrsim10$ around important spectral features for identifying the galaxy redshifts. We used a central wavelength of 6350~\AA\ for the first exposure and applied offsets of 100~\AA\ toward the blue and the red for the following exposures to avoid the loss of any spectral features that could, by chance, lie in the gaps between CCDs. For both masks, spectroscopic flats and spectra of CuAr comparison lamps were taken before or after each science exposure. The spectrophotometric standard star LTT 7379 was observed with the same instrument setup but on a different night (2024 October 21 UT) and under different observing conditions. Therefore, only a relative flux calibration was provided.

 
All spectra were reduced using the semi-automated Python Spectroscopic Data Reduction Pipeline \citep[Pypeit;][]{Prochaska_Pypeit_2020}. The science exposures, spectroscopic flats, and CuAr comparison lamps were overscan-subtracted, bias-subtracted, and trimmed; additionally, cosmic rays were removed. 
For each wavelength setting, the normalized spectroscopic 
flat was used to automatically find the slits, trace the edges, and flat-field the two-dimensional science spectra. 
The arc frames, plus an archival GMOS-S wavelength solution, were used to generate a map of the wavelength solution across the entire detector for each slit. Using mask design information, Pypeit jointly performed object extraction and b-spline sky subtraction and corrected for spectral flexure with reference to the sky emission lines. 
The three 2D science exposures for each MOS mask were then co-added and the spectra extracted to 1D format. Finally, the extracted spectra were calibrated in flux using our spectrophotometric standard and corrected for atmospheric extinction using an archival extinction curve from the nearby Cerro Tololo Inter-American Observatory.
The obtained $rms$ for the wavelength solution varied between $\sim$0.17 and $\sim$0.30 pixels ($\sim$0.2 \AA~to $\sim$0.38 \AA). The final spectra have a resolution of $\sim$6.5 \AA~(from the FWHM of the sky lines), a wavelength dispersion of $\sim$1.26 \AA~ pixels $^{-1}$, and an average wavelength coverage of $\sim$4200-8100 \AA, although the coverage depends on the position of the slit in the GMOS mask. 


\subsection{Redshift Calculation}
\label{subsec:redshifts}

We determined the redshifts of the observed galaxies in the heliocentric reference frame using two programs implemented inside the IRAF RV package. The RVIDLINES package employs a line-by-line Gaussian fit for spectra with emission features. The shifts of the fitted lines are then averaged to produce a final measurement of the radial velocity, while the error is the standard deviation of the line shifts \citep[e.g.,][]{Kaldare_RVIDLINES_2003}. The FXCOR program implements the cross-correlation algorithm \citep{Tonry_R_1979}, with estimated uncertainties based on the statistic value $R$, although we note that these derived uncertainties are typically $\sim 1.6$ times smaller than the true uncertainties \citep{Quintana2000}. 
After sparingly removing obvious sky emission line subtraction residuals and bad pixels, we cross-correlated each early-type spectrum with four high S/N templates and chose the redshift solution with the highest $R$ ($R \geqslant 5$) that was consistent within uncertainties with the result obtained from at least one other template. Overall, we measured the redshifts for 76 of the 77 galaxies targeted in the two masks ($\sim$99$\%$ success rate).
\begin{deluxetable*}{lccccccccc} 
\digitalasset
\tablewidth{0 pt} 
\tablecaption{Galaxy Redshift Catalog}
\tablehead{
\colhead{Galaxy} & \colhead{R.A.} & \colhead{Decl.} & \colhead{$r$} & \colhead{$g-r$} & \colhead{$z \pm \Delta z$} & \colhead{$R$ / \#L} & \colhead{$N_Z$} & \colhead{Member?} & \colhead{Subcluster} \\
\colhead{(1)} & \colhead{(2)} & \colhead{(3)} & \colhead{(4)} & \colhead{(5)} & \colhead{(6)} & \colhead{(7)} & \colhead{(8)} & \colhead{(9)} & \colhead{(10)}
}
\startdata 
J025902.83-193853.1 & 44.76183 & -19.64811 & 19.91 & 1.48 & $0.269921 \pm 0.000144$ & 7.2 / -- & 2 & Yes & A \\ 
J025900.13-193937.5 & 44.75058 & -19.66044 & 20.64 & 1.05 & $0.268442 \pm 0.000081$ & -- / 10 & 1 & Yes & B \\ 
J025907.68-193642.8 & 44.78204 & -19.6119 & 20.82 & 1.2 & $0.266686 \pm 0.000133$ & 12.8 / -- & 1 & Yes & C \\ 
J025908.90-193729.0\tablenotemark{a} & 44.78709 & -19.62474 & 19.19 & 1.32 & $0.249405 \pm 0.000127$ & 13.2 / -- & 1 & No &  \\ 
J025905.05-193924.5 & 44.77105 & -19.65682 & 20.81 & 1.27 & $0.282189 \pm 0.000066$ & 6.3 / 11 & 2 & No &  \\
\enddata
\tablecomments{Columns: 
(1) galaxy name; 
(2 and 3) J2000 right ascension (R.A.) and declination (decl.), both in decimal degrees; 
(4 and 5) total LSDR10  dereddened $r$ magnitudes and $g-r$ color; 
(6) heliocentric redshift and uncertainty; 
(7) $R$-values for results of the cross-correlation technique from \citet{Tonry_R_1979} and/or the number of emission lines used to calculate redshift;
(8) number of redshift measurements averaged for the final adopted redshift; 
(9) membership flag;
(10) assigned subcluster. 
Table \ref{tbl:catalog} is published in its entirety in the machine-readable format. A portion is shown here for guidance regarding its form and content.
\label{tbl:catalog}
}
\tablenotetext{a}{Galaxy member of a foreground group. See Fig. 4 and section \S~\ref{subsec:LOS_substructure} for details.}
\end{deluxetable*}

Of the 76 galaxies with measured redshifts, six were observed twice (once through each mask). 
As expected, we do not find evidence for a systematic offset between the measured redshifts from the two masks, giving us confidence in our determination of the redshift.
For these galaxies, we averaged the measurements and calculated the redshift uncertainty following Equation 2 of \citet{Quintana2000}. A table of the resulting 76 heliocentric redshifts and corresponding uncertainties is available online; a few sample rows are shown in Table~\ref{tbl:catalog}.

\subsection{Completeness of the Spectroscopic Sample}
\label{subsec:completeness}

\begin{deluxetable}{lccccc} 
\tablecaption{RCS Spectroscopic Completeness Fractions 
\label{tbl:complete}}
\tablehead{
\colhead{$r$ Limit} & \multicolumn{5}{c}{Maximum Radius} \\
\colhead{} & \colhead{1.0'} & \colhead{2.0'} & \colhead{3.0'} & \colhead{4.0'} & \colhead{5.0'} \\
\colhead{(1)} & \colhead{(2)} & \colhead{(3)} & \colhead{(4)} & \colhead{(5)} & \colhead{(6)} 
}
\startdata 
$r < 18.0$ & 1.0 & 1.0 & 0.80 & 0.86 & 0.86 \\ 
$r < 19.0$ & 1.0 & 1.0 & 0.88 & 0.78 & 0.75 \\ 
$r < 20.0$ & 0.67 & 0.71 & 0.74 & 0.62 & 0.56 \\ 
$r < 21.0$ & 0.46 & 0.54 & 0.61 & 0.55 & 0.48 \\
\enddata
\tablecomments{Cumulative spectroscopic completeness fractions out to five maximum radii from the cluster's X--ray peak and to four $r$-band apparent magnitude limits. We used the LSDR10 dereddened magnitudes. All fractions are calculated only considering the galaxies selected photometrically to be RCS members.}
\end{deluxetable} 

Here we determine the completeness of our spectroscopic sample to various magnitude limits and maximum projected clustercentric radii. We obtained spectra for galaxies as far as $\sim$5.0\arcmin\ from the cluster's X--ray peak.
Therefore, we adopt a maximum radius of $5.0'$ when calculating the completeness fractions. Each completeness fraction is defined as the number of galaxies with measured redshifts that satisfy the specified radius and magnitude limits divided by the total number of galaxies from LSDR10 satisfying the same limits.

We achieve $\sim$34$\%$ completeness for $r<21.0$ inside a maximum radius of $5.0'$ from the X--ray peak. For reference, at $z \approx 0.267$, these limits correspond to an absolute $r$-band magnitude of  $M_r\cong-20.0$\footnote{We apply a $K$-correction using the median dereddened LSDR10 $griz$ fluxes and uncertainties for surveyed galaxies via the \texttt{kcorrect} software package \citep{kcorrect}.} and a limiting radius of $0.56 R_{200}$ (see section \ref{subsec:zdist}).
Since we selected galaxies for observation from the RCS (see \S~\ref{subsec:selecttargets}), we also calculated the completeness fractions while only considering RCS galaxies (see Table \ref{tbl:complete}).
Among RCS galaxies, we achieve a $\sim$48$\%$ completeness for $r<21.0$ within a radius of $5.0'$, giving us confidence that our spectroscopic sample represents the cluster's galaxy population well.
\section{Cluster Dynamical Parameters}
\label{sec:analysis}


\subsection{Redshift Distribution}
\label{subsec:zdist}

\begin{figure}
    \centering
    \includegraphics[width=1.0\linewidth]{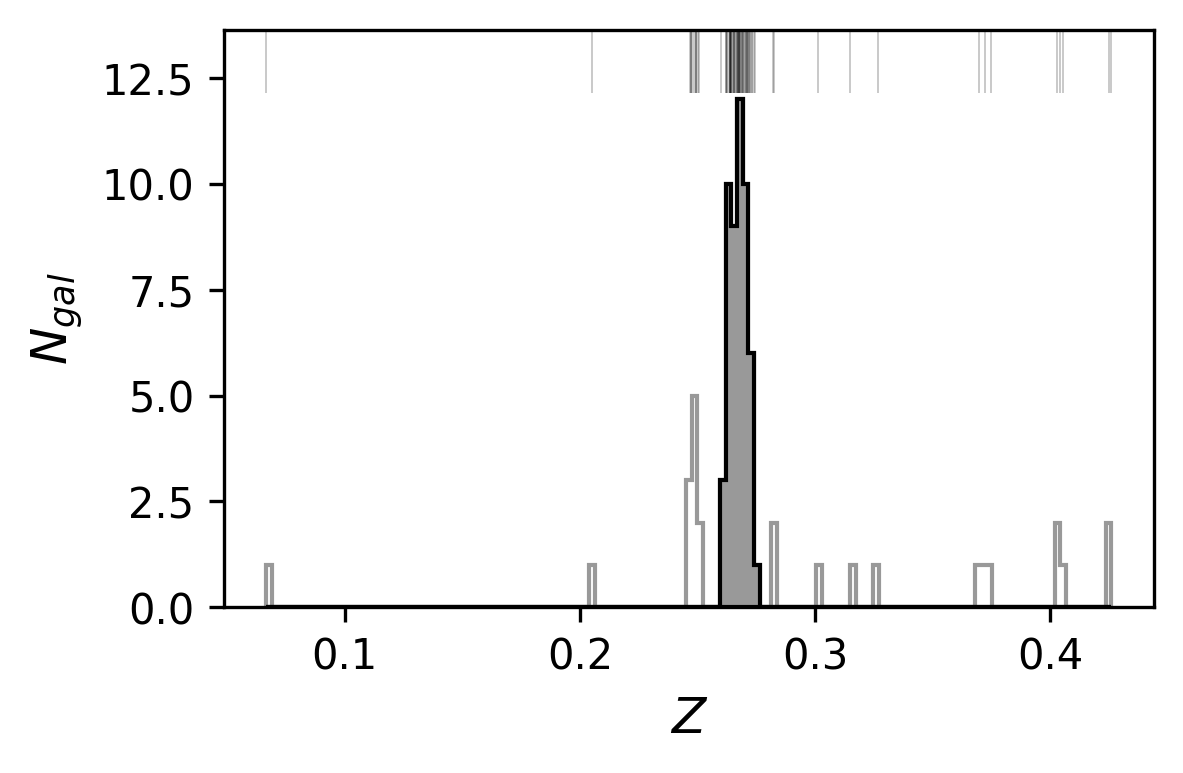}
    \caption{Redshift distribution of the 76 galaxies with secure spectroscopic redshifts (bin width of $\Delta z \cong 0.0024$). The shaded gray region indicates member galaxies at the redshift of the cluster ($\overline{z} \approx 0.267$).
    The tick marks at the top of the figure represent the redshifts of individual galaxies.}
    \label{fig:z_distribution}
\end{figure}

Fig. \ref{fig:z_distribution} shows the histogram of the redshift distribution for the 76 galaxies in the field of J025859. To determine the member galaxies,
we first identify the highest peak in the redshift distribution (center of the tallest histogram bin at $z \approx 0.264$).
We use this approximate cluster redshift to obtain a preliminary value for each galaxy's peculiar velocity 
\begin{equation}\label{eq:vpec}
    V_{\rm{pec}} = c \frac{z - \overline{z}}{1 + \overline{z}},
\end{equation}
\noindent which we use to select all galaxies with $|V_{\rm{pec}}|\leqslant 5000~\rm{km~ s^{-1}}$. From this subset of galaxies, an iterative $3 \sigma$ clipping algorithm \citep{Yahil1977} is applied to isolate members of the galaxy cluster and remove background/foreground interlopers. The robust bi-weight estimators of the central location and scale of \citet{Beers_location_1990} are used to estimate the cluster redshift $\overline{z}$
and the observed 1D velocity dispersion of the cluster along the line of sight $\sigma_{\rm{obs}}$. The procedure is repeated until the number of member galaxies converges to a constant value (51 cluster members in 4 iterations). The selected member galaxies are highlighted in the observed redshift distribution in Fig. \ref{fig:z_distribution} and in the CMD in Fig. \ref{fig:member_CMD}. 
As shown in the CMD, the vast majority of member galaxies are located inside the RCS region, which is expected since the RCS was used to prioritize member candidates for spectroscopic follow-up (see \S~\ref{subsec:selecttargets}).
The observed line-of-sight velocity dispersion is then corrected to the cluster reference frame as follows:

\begin{equation}
     \sigma_{\rm{cl}} = \frac{\sigma_{\rm{obs}}}{1 + \overline{z}}.
\label{eq:sigcl}
\end{equation}

Finally, we apply to our calculated $\sigma_{\rm{cl}}$ the small \citet{Ferragamo2020} corrections, which take into account the statistical and physical effects in small samples of cluster members. In particular, the unbiased estimator of $\sigma_{\rm{cl}}$ is calibrated based on the number of member galaxies used for the calculation. 
We apply a further aperture-subsampling correction to our $\sigma_{\rm{cl}}$ value, which is based on $R_{\rm{max,opt}}/R_{200}$. Here, $R_{\rm{max,opt}}$ is defined as the maximum projected radius of a confirmed cluster member from the optical center of the cluster, which was identified by \citetalias{Kluge_optical_2024} and lies very close to the X--ray peak ($\sim$10.7\arcsec\ separation 
between the optical center and X--ray peak, see \S~\ref{subsec:2d_substructure}), and $R_{200}$ is calculated following \citet{Carlberg_mass_1997} as:
\begin{equation}
    R_{200} = \frac{\sqrt{3} \sigma_{\rm{cl}}}{10 H(\overline{z})}.
    \label{eq:r200}
\end{equation}
We apply the corresponding correction factor using a linear interpolation of the factors provided in Table 4 of \citet{Ferragamo2020}. 

Table \ref{tbl:properties} shows the resulting corrected $\sigma_{\rm{cl}}$, along with $\overline{z}$, the number of member galaxies ($N_{\rm{mem}}$), 
$R_{\rm{max,opt}}$, $R_{200}$, 
and other relevant dynamical parameters derived in the following subsections. 


\begin{deluxetable}{@{\hspace{0.09\textwidth}}lc@{\hspace{0.09\textwidth}}}
\tablecaption{Cluster Properties}
\tablehead{
\colhead{Parameter} & \colhead{Quantity} \\
\colhead{(1)} & \colhead{(2)} 
}
\startdata 
R.A. (2000) & $02^{\rm h}58^{\rm m}59^{\rm s}.40$ \\ 
Decl. (2000) & $-19^{\circ}37^{\prime}27^{\prime\prime}.36$ \\ 
$N_{\rm{mem}}$ & $51$ \\ 
$\overline{z}$ & $0.26744^{+0.00049}_{-0.00057}$ \\ 
$\sigma_{\rm{cl}}~\rm{(km~s^{-1})}$ & $808^{+59}_{-66}$ \\ 
$D_{\rm{BCG - X}}~\rm{(Mpc)}$\tablenotemark{a} & $0.954$ \\ 
$R_{\rm{max,opt}}~\rm{(Mpc)}$ & $0.921$ \\ 
$R_{200}~\rm{(Mpc)}$ & $1.79^{+0.13}_{-0.15}$ \\ 
$M_{200}$ ($10^{14}~M_{\odot}$) & $4.47^{+0.94}_{-0.93}$ \\ 
\enddata
\tablecomments{  
All uncertainties are at the $1\sigma$ level and were obtained via a bootstrapping approach with $10^5$ realizations. The cluster coordinates are from the eRASS1 galaxy groups and clusters main catalog \citep{Bulbul_clusters_2024}. 
\tablenotetext{a}{Distance to the BCG determined in \S~\ref{subsec:LOS_substructure}.}
\label{tbl:properties}
} 
\end{deluxetable} 

\subsection{Cluster Mass Estimation}
\label{subsec:clustermass}



We compute the dynamical mass of the cluster using the unbiased mass estimator by \citet{Ferragamo2020}, which uses the corrected $\sigma_{\rm{cl}}$ as an input (see \S~\ref{subsec:zdist}). The mass estimator employs the $\sigma_{\rm{cl}}$-$M_{200}$ scaling relation of \citet{Munari_mass_2013}, which is derived from zoomed-in hydrodynamical simulations of galaxies within DM halos and takes into account prescriptions for cooling, star formation, supernova feedback, and feedback from active galactic nuclei:
\begin{equation}
\sigma_{\rm{cl}} = A_{\rm{1D}} \left[\frac{h(\overline{z}) M_{200}}{10^{15} M_{\odot}}\right]^\alpha,
\end{equation}
where 
$A_{\rm{1D}} = 1177 \pm 4.2~ \rm{km ~s^{-1}}$, $\alpha = 0.364 \pm 0.0021$, $h(\overline{z}) = H(\overline{z})/(100~\rm{km~s^{-1}~Mpc^{-1}})$, and $M_{200}$ is the mass within $R_{200}$.
The resulting value for $M_{200}$ is listed in Table \ref{tbl:properties}.


Our dynamically derived mass estimate $M_{200} = 4.47^{+0.94}_{-0.93} \times10^{14}~M_{\odot}$ is consistent within its $1\sigma$ uncertainties with the previous SZ-based estimate of $M_{200} = 5.22^{+1.26}_{-1.02} \times 10^{14}~M_{\odot}$ \citep{Hilton_ACT_2021}, which was estimated utilizing the \citet{Arnaud_YSZ_2010} scaling relation between SZ signal and mass, a richness-based weak-lensing mass calibration factor, and the \citet{Bhattacharya_DM_2013} $c$-$M$ relation. 
However, we caution that the precision of our mass estimate is limited by our mapping of the cluster members to $\sim$0.5$R_{200}$ (see Table \ref{tbl:properties}) and the precision of the subsequent aperture-subsampling correction we apply \citep{Ferragamo2020}. Additional uncertainty is introduced by the fact that the \citet{Munari_mass_2013} relations are calibrated on isolated clusters, without considering cluster mergers \citep[e.g.,][]{Monteiro-Oliveira22-Hercules}.



 
 


\section{Substructure Analysis} 
\label{sec:substructure}

In this section, we analyze the dynamical state of J025859 by using one- and two-dimensional statistical tools to examine the presence of substructures along the line of sight and in the projected distribution of member galaxies in the plane of the sky. Additionally, we compare the location of each identified substructure with the cluster's X--ray morphology and examine the substructures' locations in projected phase space.

\subsection{Line-of-sight Substructures}
\label{subsec:LOS_substructure}


\begin{figure*}
    \centering
    \includegraphics[width=1.0\linewidth]{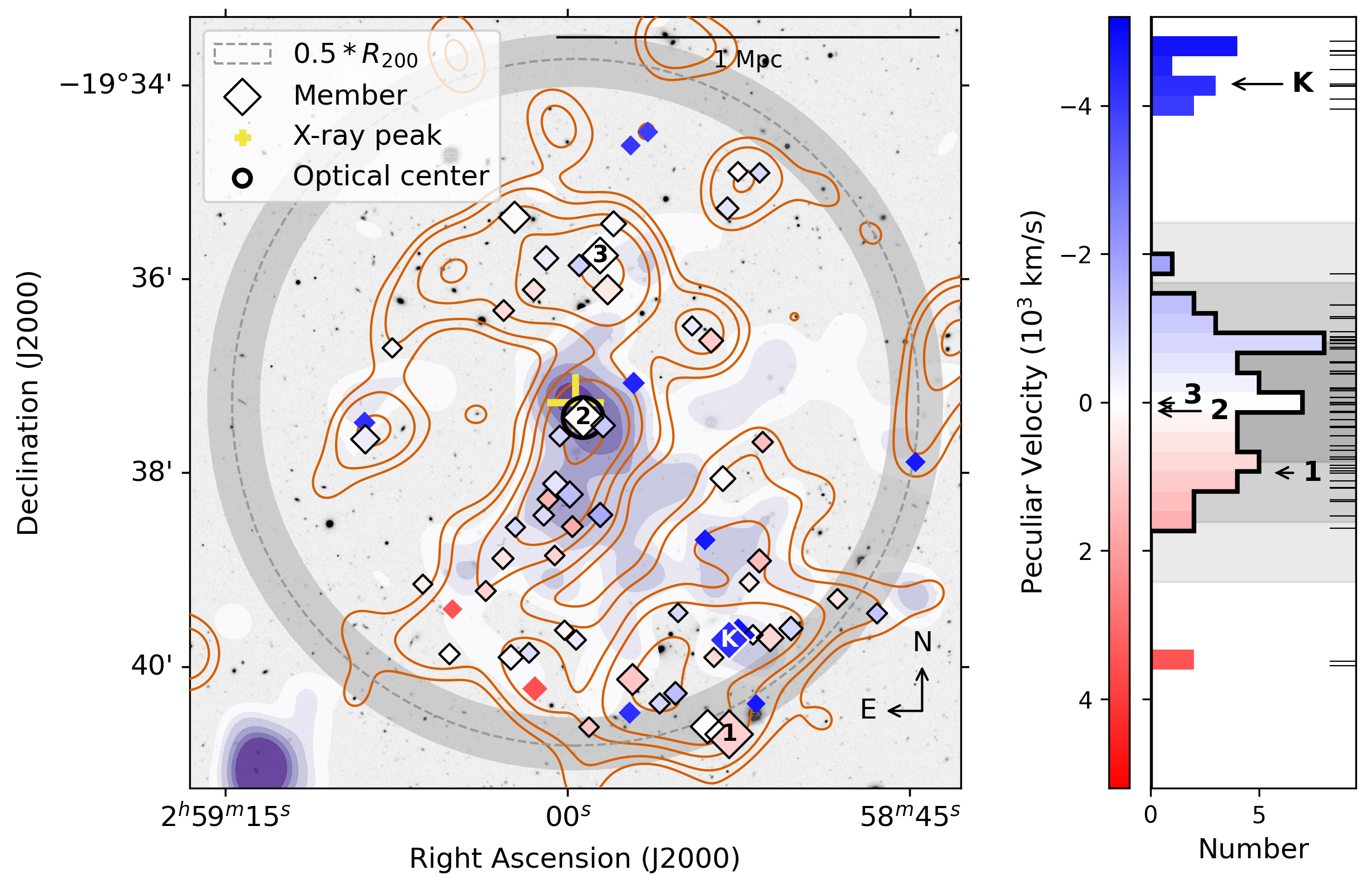}
    \caption{
    Left panel: Projected cluster distribution. The 63 galaxies with estimated peculiar velocities within $\pm$5000 km s$^{-1}$ of the cluster rest frame are shown as colored rhombuses; black solid borders indicate cluster members. The color of the rhombuses denotes the peculiar velocity. 
    The size of each rhombus is proportional to the galaxy stellar mass; the three brightest spectroscopic member galaxies are labeled with the numbers 1-3. The optically-selected BCG \citepalias{Kluge_optical_2024} is labeled with \enquote{K}. The X--ray peak and optical center \citepalias{Kluge_optical_2024} are shown by a yellow + and a black O, respectively. The large gray dashed circle and the shaded gray annulus show the $0.5R_{200}$ and the $1 \sigma$ uncertainty, respectively. The purple contours correspond to the eROSITA X--ray surface brightness (same as Fig. \ref{fig:cluster_inspector}).
    The red contours show the kernel density map of the RCS galaxies. In the background is the $r$-band LSDR10 image of the cluster.
    Right panel: peculiar velocity distribution of the spectroscopically observed galaxies. The black line represents member galaxies. The gray background shading represents one, two, and three velocity dispersions. The tick marks at the right border of the plot represent individual peculiar velocities. 
    }
    \label{fig:density_and_histogram}
\end{figure*}

The right panel of Fig. \ref{fig:density_and_histogram} shows the updated peculiar velocity distribution, recalculated using $\overline{z}$ and $\sigma_{\rm{cl}}$ (see \S~\ref{subsec:zdist} and Table \ref{tbl:properties}) for the member galaxies (solid black line). 
Although the histogram suggests the presence of subgroups of member galaxies 
within 3$\sigma$, we are unable to reject the null hypothesis of a single Gaussian distribution with the Anderson-Darling test ($p\sim0.55$), Shapiro-Wilk test ($p\sim0.53$), or Hartigan's dip test for unimodality ($p \sim 0.66$). 



In addition to the member galaxies, the right panel of Fig. \ref{fig:density_and_histogram} reveals the presence of a \enquote{foreground group} of 10 galaxies at $z \sim 0.249$ (see the discussion in \S~\ref{subsec:foreground}).
This foreground group includes the BCG selected photometrically by \citetalias{Kluge_optical_2024}, revealing that this candidate BCG is not a member of the main cluster. We identify the correct BCG as J025852.92-194041.9 (marked \enquote{1} in Fig.~\ref{fig:density_and_histogram}), which is the brightest spectroscopically confirmed cluster member and the brightest LSDR10 galaxy within $\sim 0.5\ R_{200}$. 
The BCG underscores the disturbed state of the cluster, as it displays a large offset from the X--ray peak $D_{\rm{BCG - X}}= 0.954~{\rm{Mpc}}$ (see Table \ref{tbl:properties}). This large offset is explained by the BCG's membership in a recently accreted subcluster (see \S~\ref{subsec:phase_space}).

\subsection{Projected Galaxy Distribution}
\label{subsec:2d_substructure}

The left panel of Fig. \ref{fig:density_and_histogram} shows the projected distribution of the cluster members (rhombuses with black borders), a Gaussian kernel density map \citep{Silverman_KDE_1998} of the optically-selected RCS member candidates (red contours), and the eROSITA X--ray surface brightness contours (see figure caption for details).
From Fig.~\ref{fig:density_and_histogram}, it is clear that the distribution of the confirmed member galaxies closely follows the RCS galaxy density contours, which reflects our selection of spectroscopic targets 
(see \S~\ref{subsec:selecttargets}). Additionally, the member and RCS galaxy distributions display at least three overdensities: the central subcluster and two other substructures $\sim$2\arcmin \ northeast and $\sim$4\arcmin \ southwest of the optical center of the cluster, respectively. Interestingly, each of the three overdensities aligns with one of the cluster's three BCGs (marked 1, 2, and 3). We present more information about each BCG, including coordinates, magnitudes, and peculiar velocities, in Table \ref{tbl:BCGs}.

\begin{deluxetable}{lccc} 
\tablewidth{0 pt} 
\tablecaption{BCG Parameters}
\tablehead{
\colhead{BCG} & \colhead{Subcluster} & \colhead{$M_{r}$} & \colhead{$V_{\rm{pec}} \pm \Delta V_{\rm{pec}}$} \\
\colhead{(1)} & \colhead{(2)} & \colhead{(3)} & \colhead{(4)} \\
}
\startdata 
J025859.31-193725.9 & A & -23.63 & $118 \pm 150$ \\ 
J025852.92-194041.9 & B & -23.97 & $951 \pm 156$ \\ 
J025858.58-193545.6 & C & -23.34 & $12 \pm 145$ \\ 
J025852.92-193943.6 & K24 & -23.55 & $-4291 \pm 147$ \\
\enddata
\tablecomments{  
Columns: 
(1) galaxy name; 
(2) Corresponding substructure (see subsection \ref{subsec:2d_substructure});
(3) total absolute $r$ magnitude, corrected for reddening and $K$-corrected using LSDR10 $griz$ photometry via \texttt{kcorrect} \citep{kcorrect};
(4) peculiar velocity with respect to the cluster rest frame (see Equation \ref{eq:vpec}) and associated uncertainty calculated using the quadrature method. The BCG of the foreground group (J025852.92-193943.6) was selected photometrically as the cluster BCG by \citetalias{Kluge_optical_2024}.
\label{tbl:BCGs}}
\end{deluxetable}


To further investigate these apparent overdensities, we input the R.A. and decl.
of the member galaxies into \texttt{mclust}\footnote{cran.r-project.org/web/packages/mclust/index.html} \citep{mclust}, an R package for Gaussian mixture models with several past applications in substructure analysis \citep[see, e.g.,][]{
Monteiro-Oliveira_A1644_2020, Kim_merging_2024}. \texttt{mclust} uses an expectation maximization algorithm for normal mixture models \citep{Scrucca_Mclust_2016} to fit the input multidimensional data with Gaussian components using various covariance structures. To determine the optimal number and structure of components, \texttt{mclust} returns two different criteria: the Bayesian information criterion (BIC) and the integrated complete-data likelihood (ICL) criterion \citep{Biernacki_ICL_2000}. After fitting models with between one and nine components with 14 covariance structure options, we select the only model that appears within the top three scores for both BIC and ICL; this model divides the galaxy distribution into three components with the covariance structure \enquote{EEV} (same shape, same volume, varying orientation).

\begin{figure}
    \centering
    \includegraphics[width=1.0\linewidth]{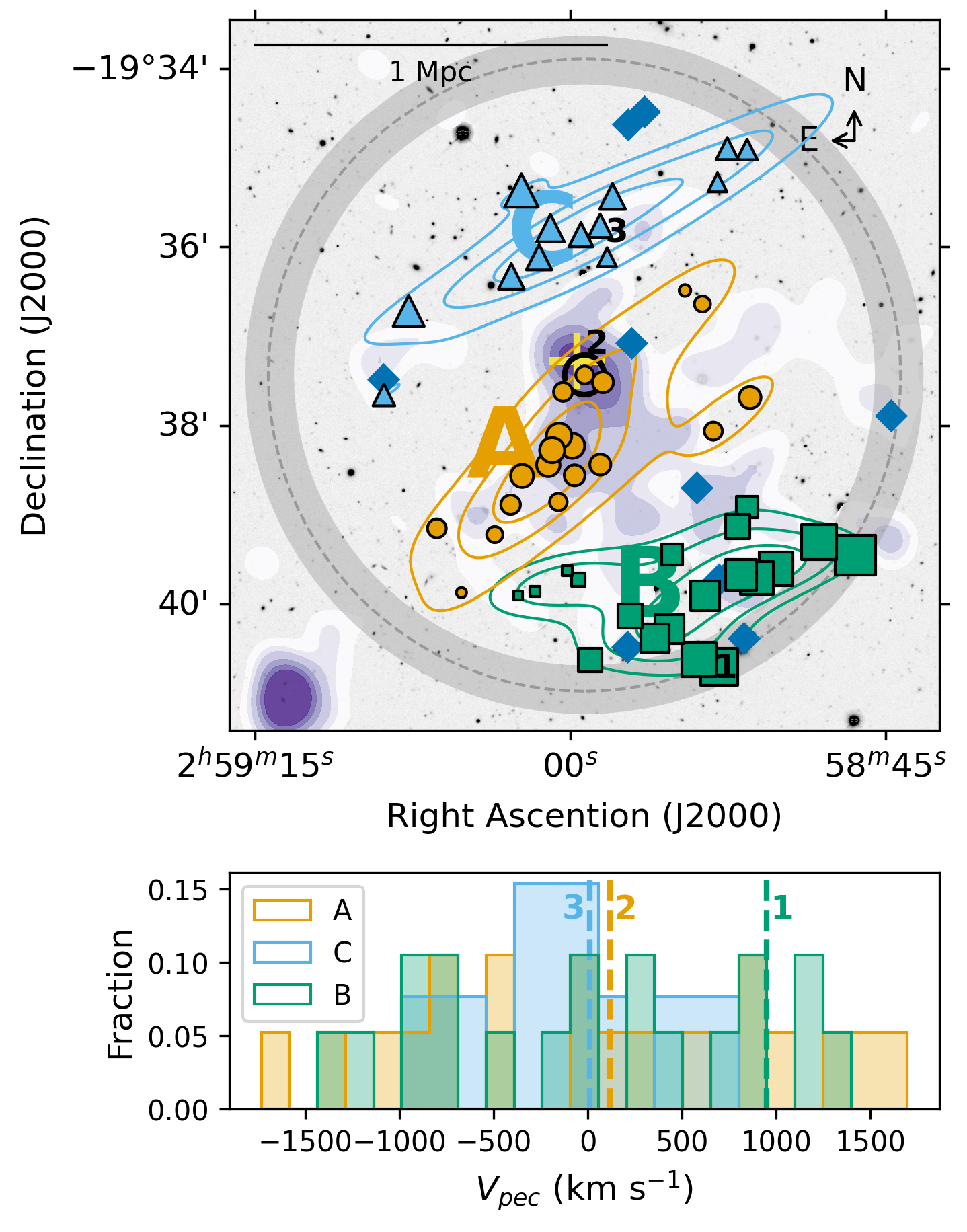}
    \caption{
    Top panel: the three subclusters identified in the 2D cluster density map using the \texttt{mclust} algorithm. Galaxies associated with \texttt{mclust} groups A (central), B (southwestern), and C (northeastern) are shown by orange circles, green squares, and light blue triangles, respectively, and surrounded by their kernel density map contours in the same color. A larger shape indicates that the galaxy was classified into a subcluster with lower classification uncertainty.
    The dark blue rhombuses show galaxies located in the \enquote{foregound group}.
    The cluster image, BCGs, eROSITA X--ray contours, $0.5R_{200}$, X--ray peak, and optical center are the same as in Fig. \ref{fig:density_and_histogram}.
    Bottom panel: peculiar velocity (Equation \ref{eq:vpec}) distribution of each subcluster in the cluster rest frame. The vertical lines represent the BCGs (see Table \ref{tbl:BCGs}).
    }
    \label{fig:mclust_groups}
\end{figure}

The top panel of Fig. \ref{fig:mclust_groups} shows the location of the galaxies associated with each identified subcomponent, which are clearly coincident with the overdensities visible in Fig. \ref{fig:density_and_histogram}. For each group, Table \ref{tbl:mclust} lists basic properties, including 
$N_{\rm{mem}}$; $\overline{z}$, calculated using the biweight location; and $\sigma_{\rm{cl}}$ (Equation \ref{eq:sigcl}), for which $\sigma_{\rm{obs}}$ was estimated using the biweight scale statistic for $N_{\rm{mem}} > 15$ (subclusters A \& B) and otherwise using the preferred gapper estimator \citep{Ruel_SPT_2014, Beers_location_1990, Wainer_regression_1976}. In addition, the same measures for the \enquote{foreground group} of galaxies are presented.
We again apply the \citet{Ferragamo2020} corrections when calculating $\sigma_{\rm{cl}}$ to mitigate the effects of statistical bias and aperture subsampling. For the aperture subsampling correction, we used the biweight location of the coordinates of each subcluster's assigned galaxies as a centroid from which to measure $R_{\rm{max}}$; these centroids are listed as subcluster coordinates in Table \ref{tbl:mclust}.

\begin{deluxetable*}{lccccccc} 
\tablewidth{0 pt} 
\tablecaption{Subluster Properties}
\tablehead{
\colhead{Structure} & \colhead{R.A.} & \colhead{Decl.} & \colhead{$N_{\rm{mem}}$} & \colhead{$\overline{z}$} & \colhead{$\sigma_{\rm{cl}}$ (km s$^{-1}$)} \\ 
\colhead{(1)} & \colhead{(2)} & \colhead{(3)} & \colhead{(4)} & \colhead{(5)} & \colhead{(6)} \\ 
}
\startdata 
A & $02^{\rm h}59^{\rm m}00^{\rm s}.30$ & $-19^{\circ}38^{\prime}15^{\prime\prime}.06$ & 19 & $0.2676^{+0.0011}_{-0.0014}$ & $980^{+110}_{-140}$ \\ 
B & $02^{\rm h}58^{\rm m}54^{\rm s}.36$ & $-19^{\circ}39^{\prime}47^{\prime\prime}.30$ & 19 & $0.26765^{+0.00100}_{-0.00094}$ & $805^{+86}_{-100}$ \\ 
C & $02^{\rm h}58^{\rm m}59^{\rm s}.53$ & $-19^{\circ}35^{\prime}45^{\prime\prime}.54$ & 13 & $0.26685^{+0.00064}_{-0.00063}$ & $526^{+43}_{-124}$ \\ 
Foreground & $02^{\rm h}58^{\rm m}54^{\rm s}.80$ & $-19^{\circ}38^{\prime}14^{\prime\prime}.93$ & 10 & $0.24869^{+0.00060}_{-0.00074}$ & $335^{+17}_{-83}$ \\ 
\enddata
\tablecomments{Columns: (1) subcluster identification; (2 and 3) coordinates;
(4) number of assigned galaxies; (5) redshift;
(6) velocity dispersion along the line of sight, corrected to the subcluster rest frame.
All uncertainties are at the $1\sigma$ level and were estimated using $10^5$ bootstrap realizations.
\label{tbl:mclust}} 
\end{deluxetable*} 


Overall, as illustrated in both Table \ref{tbl:mclust} and the bottom panel of Fig. \ref{fig:mclust_groups}, the \texttt{mclust}-identified groups have very similar redshifts. Groups A and B (central and southwest) are the most similar, with a velocity difference of only $|\Delta V| \sim 7~\rm{km~s^{-1}}$ in the cluster rest frame (Equation \ref{eq:vpec}), well within the $1 \sigma$ uncertainties. In comparison, the northeast group C is slightly blueshifted, with velocity offsets of $|\Delta V| \sim 190~\rm{km~s^{-1}}$ and $|\Delta V| \sim 180~\rm{km~s^{-1}}$ compared to subclusters A and B, respectively. Overall, the finding of three clear subclusters in the plane of the sky
with similar redshifts, and the absence of substructure along the line of sight, strongly suggests a cluster merger along the plane of the sky. 

Despite their consistent redshift, the subclusters vary in their number of member galaxies and velocity dispersion. Subclusters A and B are assigned more member galaxies ($N_{\rm{mem}} =19$)
and display higher velocity dispersions; subcluster A's velocity dispersion ($\sigma_{\rm{cl}} =980~\rm{km~s^{-1}}$) actually exceeds the velocity dispersion for the whole cluster ($\sigma_{\rm{cl}} =808~\rm{km~s^{-1}}$). In contrast, we survey fewer member galaxies in subcluster C ($N_{\rm{mem}} =13$), which shows a significantly lower velocity dispersion ($>2\sigma$ significance). 
As expected, each of the three subclusters contains one of the cluster's three BCGs.

\subsection{Comparing Substructure with X--ray Morphology}
\label{subsec:X--ray_substructure}

Here we compare the position of the identified substructures with the cluster's X--ray contours, as shown in Fig.~\ref{fig:mclust_groups}.
The central subcluster (A) seems the most associated with the X--ray emission. The BCG of this subcluster is selected as the optical center of the cluster and aligns quite well with the cluster's eROSITA X--ray peak, with an offset of only $\sim$10.7\arcsec\ ($\sim45~\rm{kpc}$). This offset is comparable 
to the median positional uncertainty of eROSITA of $\approx$ 4.65\arcsec\ \citep{Brunner_eFEDS_2022}. This indicates some relaxation, as both the optical center and the X--ray peak are located at a minimum of the gravitational potential well of the subcluster. However, we notice that the bulk of the galaxy members of this subcluster seem to be displaced by $\sim$0.5\arcmin\ to the southeast and aligned with a smaller X--ray lobe extending south from the X--ray peak. 

Furthermore, it appears that the southwest subcluster (B) overlaps only with the edges of X--ray emission rather than an overdensity, and the northeast subcluster (C) is not associated with any X--ray emission. This suggests the occurrence of a dissociative merger, in which the collisional cluster gas has become separated from the near-collisionless galaxies \citep[e.g., ][]{Dawson_MCMAC_2013}.
However, we note that the eRASS1 X--ray images are of limited depth, with an average exposure time of 199 seconds for this cluster \citep{bulbul24}. Thus, in order to see the full extent of the X--ray morphology and draw stronger conclusions, deeper X--ray imaging is needed, including that from upcoming eROSITA data releases.



\subsection{Projected Phase-space Diagram}
\label{subsec:phase_space}

\begin{figure}
    \centering
    \includegraphics[width=1.0\linewidth]{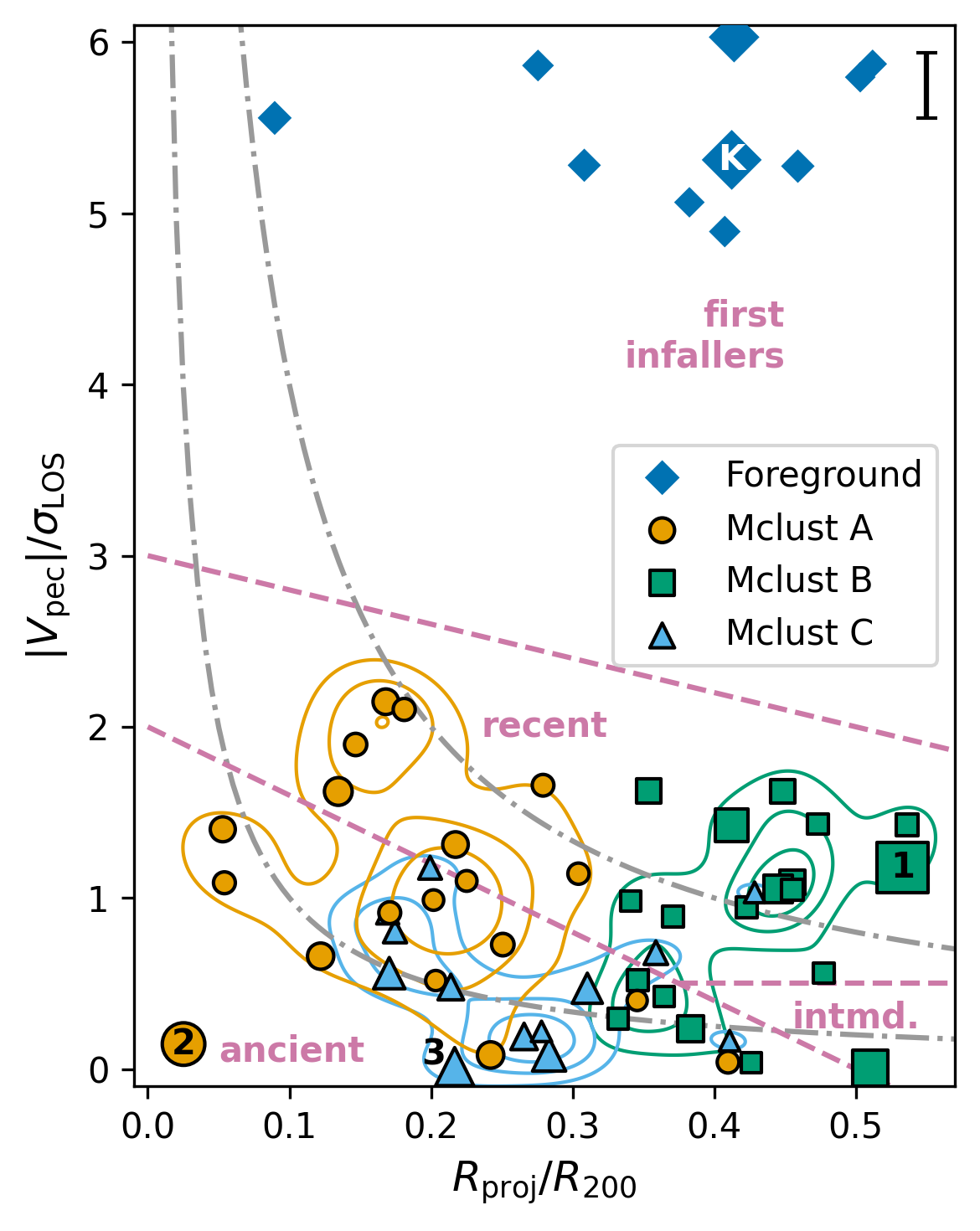}
    \caption{Projected phase-space diagram
    for each projected distribution subcluster, as well as the foreground group.
    The dash-dotted grey curves are the caustic profile lines of constant 
    $r \times v=0.1~\rm{and}~0.4$,
    following the approach of \citet{Noble_Kinematic_2013}. The pink dashed lines mark regions with different typical timescales since infall into the cluster, as proposed by \citet{Rhee_PhaseSpace_2017} and simplified by \citet{Song_A2107_2018}. Galaxies in each 
    subcluster are represented with a different shape and color, with sizes scaled off of stellar mass. BCGs are marked with numbers, as in Fig. \ref{fig:density_and_histogram}. The error bar in the upper right corner shows the median uncertainty in 
    $v$ for cluster members, calculated using the quadrature error propagation method. 
    \label{fig:phase_space_mclust}}
\end{figure}

The analysis of the projected phase-space (PPS) diagram, i.e., the examination of each member galaxy's cluster-frame velocity and radius, has proven to be a useful complementary method to investigate the dynamical state of a cluster and to differentiate between virialized cluster galaxies, i.e., those accreted recently or early on, and infalling ones \citep[see, e.g.,][]{Gill_evolution_2005, Mahajan_backsplash_2011, Haines_LoCuSS_2012, Noble_Kinematic_2013, Noble_PhaseSpace_2016, Rhee_PhaseSpace_2017, Ribero_PPS_2023}. Fig. \ref{fig:phase_space_mclust} shows the PPS diagram for all member galaxies, plus the foreground group hosting the \citetalias{Kluge_optical_2024} BCG.
We analyze the kinematics of the cluster from the PPS diagram using caustic profiles (gray dash-dot lines), i.e., lines of constant velocity–radius, following the approach described in \citet{Noble_Kinematic_2013} and the four PPS classifications introduced by \citet[][pink dashed lines]{Rhee_PhaseSpace_2017}.

Following the approach of \citet{Noble_Kinematic_2013}, 
Fig. \ref{fig:phase_space_mclust} uses two curves of constant velocity–radius
(shown as gray dash-dotted lines) to delineate three different regions in the PPS diagram:
    
    
\begin{enumerate}
    \item $0 < r \times v < 0.1$: the central region, where virialized galaxies are preferentially located;

    \item $0.1 < r \times v < 0.4$: the intermediate region, where recently accreted galaxies or the \enquote{backsplash} population of accreted, rebounding galaxies reside; and 
    
    \item $r \times v > 0.4$: the region consisting of infalling galaxies or groups,
\end{enumerate}
where 
\begin{equation}
r=\frac{R_{\rm proj}}{R_{200}},
\qquad
v=\frac{|V_{\rm pec}|}{\sigma_{\rm{cl}}},
\label{eq:rv}
\end{equation}
$\sigma_{\rm{cl}}$ is the velocity dispersion of the 
whole cluster, and $R_{\rm proj}$ is the projected radius of the member galaxy from the center of the cluster, which we define as the X--ray peak. Notably, the spectroscopically confirmed BCG (marked with \enquote{1}) is far from the gas/X--ray center, suggesting that the cluster is in a disturbed state \citep[e.g.,][]{zenteno20, Zenteno_BCG_2025}.

To associate the PPS with more specific accretion timescales, we show a complementary analysis in the pink dashed lines in
Fig. \ref{fig:phase_space_mclust}. Following the approach of \citet{Rhee_PhaseSpace_2017}, simplified by \citet{Song_A2107_2018}, we divide the member galaxies in each subcluster based on their loci in the PPS into three subsamples, each of which is associated in a statistical sense with a different range of time-since-infall ($T_{\rm{inf}}$) values:
\begin{enumerate}
    \item \enquote{ancient infallers} (associated with a $T_{\rm{inf}} \gtrsim  6.45$ Gyr),
    \item \enquote{intermediate infallers} ($T_{\rm{inf}}\sim3.6-6.45$ Gyr),
    \item \enquote{recent infallers} ($T_{\rm{inf}}\sim0-3.63$ Gyr), and
    \item \enquote{first infallers} (not infallen yet).
\end{enumerate}
Although each region contains galaxies with a broad distribution of $T_{\rm{inf}}$, the mean $T_{\rm{inf}}$ changes systematically between the different regions.


Using both schemas, we examine the location of each subcluster in the PPS diagram. Members of both the central subcluster (A, orange circles) and northeast subcluster (C, blue triangles) are spread between the intermediate and virialized  caustic 
regions and dominated by \enquote{ancient} infallers. This suggests that most galaxies were accreted during the formation of the cluster \citep[e.g.,][]{Noble_Kinematic_2013}, while some members were accreted more recently and then rebounded outward \citep[e.g.,][]{Gill_evolution_2005, Rhee_PhaseSpace_2017}. In contrast, the southwest group (B, green squares) is dominated by \enquote{recent} infallers that occupy the intermediate and infalling caustic
regions. In sum, both methods suggest a dynamically active cluster environment where the cluster is actively accreting galaxies, particularly from subcluster B.

\section{Collision Scenario Modeling}
\label{sec:collision}

With the substructures found in \S~\ref{sec:substructure}, 
we probe the merging history of the cluster using the Monte Carlo Merger Analysis Code \citep[\texttt{MCMAC},][]{Dawson_MCMAC_2013}, which was made specifically to investigate dissociative mergers. We argue this assumption is reasonable because the cluster X--ray morphology is suggestive of a dissociative merger (see \S~\ref{subsec:X--ray_substructure}). The model assumes total energy conservation and an impact parameter of zero during a collision of two spherically symmetric halos with Navarro-Frenk-White profiles \citep{Navarro_Halos_1996, Navarro_Universal_1997}. Compared to computationally-intensive $N$-body simulations and simple analytical models, \texttt{MCMAC} represents a compromise between computational time and accuracy, and it allows for further reduction of uncertainties via the addition of priors ex post facto \citep{Dawson_MCMAC_2013}.

The model takes as input the masses of the two subclusters, their redshifts, and their projected separation (as well as the associated uncertainties). \texttt{MCMAC} randomly draws values from these input probability density functions and outputs the kinematic parameters of the merger. In particular, \texttt{MCMAC} outputs the angle of the merger axis $\alpha$, where $\alpha=0\degree(90\degree)$ represents a merger in the plane in the sky (along the line of sight); the estimated distances and velocities between the clusters at different times; and the time since the first pericentric passage \citep[\enquote{time since collision},][]{Dawson_MCMAC_2013} for two possible current stages of the merger: outgoing after the first pericentric passage (TSC$_{\rm{out}}$) and incoming after reaching apoapsis (TSC$_{\rm{in}}$). Here we estimate the input subcluster masses and present the model results.

\subsection{Subclusters Mass Estimation}
\label{subsec:virial}


\texttt{MCMAC} requires as input the masses of the colliding subclusters. Besides scaling relations, another common way to estimate the mass of a galaxy cluster is through the virial theorem, which assumes that the galaxy cluster is in a state of dynamic stability \citep[e.g.,][]{Nascimento_Dynamical_2016}. Although the whole cluster is clearly not virialized, as revealed by our detection of the substructure in \S~\ref{subsec:2d_substructure}, the virial mass nevertheless provides a useful mass estimate for the individual subclusters, which are more likely to be closer to virialization. 
Assuming that a cluster is composed of approximately equal-mass bodies, we follow the method of \citet{Nascimento_Dynamical_2016} and \citet{White_merger_2015} employing the estimator for the virial mass:
\begin{equation}
M_{\rm{V}}=\frac{3 \pi}{G} \sigma_{\rm{cl}}^2 R_{\rm{H}},
\end{equation}\label{eq:virialmass}
where 
$R_{\rm{H}}$ is the mean harmonic radius, given by:
\begin{equation}
\frac{1}{R_{\rm{H}}}=\frac{2}{N_{\rm{mem}}(N_{\rm{mem}}-1)} \sum_{i<j}\frac{1}{R_{ij}},
\end{equation}
where $R_{ij}$ is the length of the projected separation vector between the $i$th and $j$th galaxies, calculated at the cluster redshift.
We use the mean harmonic radius because it more accurately reflects the effective radius of the gravitational potential of the galaxy members and is independent of the cluster center, allowing it to reflect the internal structure of the cluster \citep{Araya-Melo_halo_2009}. 

To validate the assumption that individual subclusters are virialized, we examine the Gaussianity of their individual peculiar velocity distributions \citep{Nascimento_Dynamical_2016}. We apply the same statistical tests used to look for the substructure of the cluster along the line of sight (see Section \ref{subsec:LOS_substructure}) to each subgroup, as well as the foreground group that includes the \citetalias{Kluge_optical_2024} BCG. For these small samples, the Anderson-Darling $A^2$ test is particularly useful because it is able to reliably identify deviations from Gaussianity for $N_{\rm{mem}}\geqslant 5$ \citep{DA86,Hou2009}. In all cases, we are unable to reject the null hypothesis of a single Gaussian distribution using the $A^2$ ($p \geqslant 0.42$) or Hartigan's diptest for unimodality ($p \geqslant 0.17$). 

For the three subclusters (A, B, and C), we obtain: 
\begin{align*}
M_{\rm{V, A}}=4.76^{+1.50}_{-0.97} \times 10^{14}~M_{\odot}, \\ 
M_{\rm{V, B}}=3.50^{+1.22}_{-0.74} \times 10^{14}~M_{\odot},~ \rm{and} \\
M_{\rm{V, C}}=1.52^{+0.45}_{-0.54} \times 10^{14}~M_{\odot},
\end{align*}
where the shown $1\sigma$ uncertainties in all quantities were calculated using bootstrapping with $10^5$ resamplings. Subclusters A and B have statistically indistinguishable $M_{\rm{V}}$, while subcluster C has a significantly lower mass ($>2 \sigma$) than the other two, which is consistent with the lower velocity dispersion of subcluster C.

We note that these values must be used with caution because they are clearly overestimated; indeed, $M_{\rm{V, A}}=4.76^{+1.50}_{-0.97}$ is consistent with our mass estimate for the whole cluster $M_{200}=4.47^{+0.94}_{-0.93} \times 10^{14}~M_{\odot}$ (see \S~\ref{subsec:clustermass}). This overestimation is expected given the disturbed state of the cluster, which can artificially inflate  $\sigma_{\rm{cl}}$-based subcluster mass estimates by up to a factor of $\sim$4 during the most acute merger phase \citep{hernandez-lang22,Monteiro-Oliveira22-Hercules}. However, \citet{Monteiro-Oliveira22-Hercules} found that such biased dynamical subcluster mass estimates still allowed MCMAC to recover true collision parameters ($\alpha$, TSC$_{\rm{out}}$, and TSC$_{\rm{in}}$) within  $1\sigma$ throughout a simulated cluster collision. Thus, we argue that our approximate $M_{\rm{V}}$ values can be considered as sufficient priors to constrain the collision scenario.
\subsection{MCMAC Results}
\label{subsec:mcmac_results}

Since \texttt{MCMAC} only accommodates two clusters, we choose to model the dynamics of subclusters A and B, which are the most well-sampled and massive subclusters (see \S~\ref{subsec:virial}). 
We input Gaussian priors for the virial mass and redshift values for both subclusters and adopt the larger of the upper/lower bootstrap uncertainties as the $\sigma$ for the prior. The merger parameters were obtained by sampling the possible results from $10^5$ Monte Carlo realizations. We consider only realizations with TSC$_{\rm{out}}$ less than the age of the universe at the cluster redshift ($\approx10.6$ Gyr at $\overline{z} \approx 0.267$). We apply this same criterion to TSC$_{\rm{in}}$ when calculating the statistics for TSC$_{\rm{in}}$.



\begin{deluxetable*}{lccc} 
\tablewidth{0 pt} 
\tablecaption{Collision Scenario from the MCMAC Code}
\tablehead{
\colhead{Parameter (units)} & \colhead{Value} & \colhead{Value ($\alpha \leqslant 45 \degree$)} & \colhead{Description} \\
\colhead{(1)} & \colhead{(2)} & \colhead{(3)} & \colhead{(4)} \\
}
\startdata 
$\alpha$ (deg) & $39^{+33}_{-25}$ & $22^{+16}_{-13}$ & Angle between merger axis and plane of the sky. \\ 
$d_{3D}$ (Mpc) & $0.68^{+0.75}_{-0.16}$ & $0.584^{+0.117}_{-0.086}$ & Observed 3D subcluster separation. \\ 
$d_{\mathrm{max}}$ (Mpc) & $0.95^{+1.11}_{-0.32}$ & $0.76^{+0.55}_{-0.18}$ & Maximum 3D subcluster separation. \\ 
$v_{3D}(t_{\mathrm{obs}})$ (km s$^{-1}$) & $500^{+860}_{-340}$ & $770^{+1070}_{-520}$ & 3D relative velocity at observed state. \\ 
$v_{3D}(t_{\mathrm{col}})$ (km s$^{-1}$) & $1790^{+750}_{-370}$ & $1540^{+590}_{-240}$ & 3D relative velocity at collision. \\ 
$v_{\rm{rad}}$ (km s$^{-1}$) & $250^{+300}_{-170}$ & $230^{+290}_{-160}$ & Relative radial velocity of subclusters. \\ 
TSC$_{\rm{out}}$ (Gyr) & $0.65^{+0.40}_{-0.33}$ & $0.50^{+0.20}_{-0.22}$ & Time since collision (\enquote{outgoing} scenario). \\ 
TSC$_{\rm{in}}$ (Gyr) & $1.34^{+1.16}_{-0.42}$ & $1.24^{+0.96}_{-0.38}$ & Time since collision (\enquote{incoming} scenario). \\
\enddata
\tablecomments{ 
The masses (see \S~\ref{subsec:virial}) and redshifts (see Table \ref{tbl:mclust}) for subclusters A and B were used as inputs for the simulation. Results were obtained using $10^5$ Monte Carlo realizations. The presented values are the biweight location of the distribution of the results. The uncertainties bound the 68\% bias-corrected confidence interval. For the values in (3), an ex-post-facto prior of $0 \degree \leqslant \alpha \leqslant 45 \degree$ was applied. 
\label{tbl:mcmac}
} 
\end{deluxetable*} 

The model results are summarized in Table \ref{tbl:mcmac}. In line with the methodology of \citet{hernandez-lang22}, we also show in Table \ref{tbl:mcmac} the simulation results with the addition of an ex-post-facto prior for the angle of the merger axis $\alpha \leqslant 45 \degree$. This prior is motivated by the
indistinguishable redshifts of the two subclusters (see Table \ref{tbl:mclust}),
which suggests a merger along the plane of the sky.
As expected, the constraint $\alpha \leqslant 45 \degree$ reduces uncertainty in several \texttt{MCMAC} results, including 3D subcluster separation, TSC$_{\rm{out}}$, and TSC$_{\rm{in}}$.
With the addition of this constraint, we obtain TSC$_{\rm{out}}=0.50^{+0.20}_{-0.22}~\rm{Gyr}$ and TSC$_{\rm{in}}=1.24^{+0.96}_{-0.38}~\rm{Gyr}$.
However, the uncertainties in our results are most likely underestimated because the true dynamics of the system are likely strongly affected by the third substructure, which our model does not include.

The \texttt{MCMAC} model does not directly attempt to distinguish between the two collision scenarios (\enquote{outgoing} and \enquote{incoming}). However, some insight into the likelihood of each scenario can be gained by calculating the fraction of realizations with a \enquote{valid} TSC$_{\rm{in}}$ that is less than the age of the universe at the cluster redshift \citep{Dawson_MCMAC_2013}. Among our Monte Carlo realizations, 96\% (97\%) have a valid TSC$_{\rm{in}}$ with (without) the additional constraint on $\alpha$, suggesting we cannot disregard the incoming scenario.
To further investigate these two scenarios, we compare the \texttt{MCMAC} findings with our PPS diagram analysis (see \S~\ref{subsec:phase_space}). In the PPS diagram, the majority of subcluster B member galaxies are located in the \enquote{recent infallers} region, which is associated with a time since infall of 0-3.63 Gyr \citep{Rhee_PhaseSpace_2017}. This is consistent with both time-since-collision values derived from the Monte Carlo model,
precluding us from distinguishing between the two scenarios. Ultimately, the fact that both times are relatively small highlights the recent and ongoing nature of this merger.

\section{Discussion}
\label{sec:discussion}

Overall, we find strong evidence suggesting that 1eRASS J025859.4-193727 is the result of a recent merger between clusters in the plane of the sky.
Here, we return to the cluster's X--ray morphology, particularly in the context of placing future constraints on the collision scenario. Additionally, we describe the unexpected finding of an additional, currently infalling group of galaxies near the cluster.

\subsection{X--ray Morphology}

One additional method to further constrain cluster merging scenarios is a dedicated X--ray analysis \citep[e.g.,][]{Dawson_MCMAC_2013, Doubrawa2020}. The X--ray morphology of J025859 supports our conclusion that it is an actively merging system. As shown previously (see, e.g., Fig. \ref{fig:mclust_groups}), the cluster exhibits clearly asymmetric emission in its smoothed eROSITA image. By eye, the X--ray image reveals a concentrated peak near the BCG of the central subcluster (A), with a smaller lobe that extends southward and is coincident with subcluster A's central galaxies. Apart from this central emission, there is a large, broad \enquote{tail} of fainter emission extending towards the southwest and overlapping partially with subcluster B. 


Here we examine this X--ray morphology more quantitatively using the results from \citet{Sanders_morphologies_2025}, who studied the morphologies of over 12,000 optically confirmed galaxy groups and clusters in the main eRASS1 sample \citep{Bulbul_clusters_2024}. Each cluster is assigned two composite disturbance scores: $D_{\rm{shape}}$, which is based on seven shape-based morphological metrics, and $D_{\rm{comb}}$, which also takes into consideration the cluster's concentration index. For both scores, a value of 0 (1) indicates a maximally relaxed (disturbed) morphology. Both combined metrics are based only on forward-modeled parameters, which avoid bias due to the instrumental point spread function. \citet{Sanders_morphologies_2025} find $D_{\rm{shape}}=0.957$ and $D_{\rm{comb}}=0.984$ for J025859, clearly indicating a strongly disturbed morphology.

Future studies of this asymmetric morphology may place key constraints on the cluster merger. For example, the morphology of J025859 seems qualitatively similar to the previously studied merging cluster systems Abell 3376 \citep{Monteiro-Oliveira17b} and SPT-CL J0307-6225 \citep{hernandez-lang22}. Both clusters show a \enquote{comet-like} X--ray shape with a concentrated head aligned with one galaxy subcluster trailed by a diffuse X--ray tail that overlaps with the other subcluster. In the case of Abell 3376, \citet{Monteiro-Oliveira17b} argue that one subcluster's lack of an associated X--ray peak indicates that the subcluster's gas was stripped off during the collision, similar to simulations in which the colliding subclusters had different central gas concentrations \citep{Machado_A3376_2013}. More recently, \citet{hernandez-lang22} compared the X--ray morphology of SPT-CL J0307-6225 to the gas distributions in hydrodynamic simulations from \citet{ZuHone_MergerCatalog_2018} to argue that
an \enquote{outgoing} merger scenario best describes the system. While beyond the scope of this paper, both examples highlight the potential of a dedicated X--ray analysis, combined with simulations, to add insight about the cluster's history.

\subsection{The Foreground Group}
\label{subsec:foreground}

One unexpected finding of our analysis is the presence of a foreground galaxy group that is coincident in the plane of the sky with J025859. We first found evidence of this group in the peculiar velocity distribution of surveyed galaxies (right panel of Fig. \ref{fig:density_and_histogram}), which reveals ten tightly-clustered galaxies with $\sigma_{\rm{cl}}=335^{+17}_{-83}$ and $z=0.24869^{+0.00060}_{-0.00074}$, suggesting they comprise a structure that is less massive than the J025859 subclusters and is located  $\sim$4400 $\rm{km~s^{-1}}$ from the rest-frame of the cluster (see Table \ref{tbl:mclust}). In particular, the brightest galaxy in this foreground group, J025852.92-193943.6, has $V_{\rm{pec}}=-4291 \pm147~\rm{km~s^{-1}}$. While selected photometrically as the cluster BCG by \citetalias{Kluge_optical_2024}, this galaxy and its surrounding group have clearly not yet completed their infall onto J025859.

This interpretation is supported by our analysis of the PPS diagram (Fig. \ref{fig:phase_space_mclust}). Whether considering the PPS diagram regions delineated by \citet{Noble_Kinematic_2013} or \citet{Rhee_PhaseSpace_2017}, the members of the foreground group are clearly classified as galaxies making their first infall onto the cluster. In particular, the \citet{Rhee_PhaseSpace_2017} classifications predict that the galaxies have not yet passed within $R_{200}$ of the cluster. 

Surprisingly, while closely associated in redshift space, these foreground galaxies show little spatial clustering in the projected galaxy distribution (see, e.g., Fig. \ref{fig:density_and_histogram}). The galaxies are distributed across the extent of J025859, suggesting they are infalling onto the cluster primarily along the line of sight. Within this wide distribution, there are two pairs of foreground galaxies: the first towards the north of the cluster, and the second containing the BCG of the foreground group. The latter is dynamically interesting itself; the foreground BCG and its partner have comparable brightnesses ($r=17.3~
\rm{and}~17.7$, respectively), a small separation of just $\sim$26 kpc ($6.4$\arcsec), and a peculiar velocity difference of $\sim$590 km s$^{-1}$ at the group rest-frame. Further spectroscopic data for galaxies in this field should shed light on the nature of this foreground structure. Overall, the active infall of these foreground galaxies corroborates our picture of J025859 as a dynamic and actively evolving system.



\section{Conclusions}
\label{sec:conclusion}

In this work, we have presented the first dynamical analysis and redshift catalog for the merging galaxy cluster candidate 1eRASS J025859.4-193727.
Our main results are as follows:
\begin{enumerate}
    \item We utilize Gemini South/GMOS spectra to measure the first spectroscopic redshifts for 76 galaxies inside a radius of $~5'$ from the cluster center ($\sim$1.3 $\rm{Mpc}$ at the cluster distance). We confirm 51 of these galaxies as cluster members (see Table \ref{tbl:catalog}) and confirm the BCG as J025852.92-194041.9.


    \item We determine the cluster's average redshift $\overline{z}=0.26744^{+0.00049}_{-0.00057}$ and line-of-sight velocity dispersion $\sigma_{\rm{cl}} =808^{+58}_{-65}$ km s$^{-1}$. 
    From these quantities, we estimate a dynamical mass of 
    $M_{200} = 4.47^{+0.94}_{-0.93} \times 10^{14}~M_{\odot}$, which is consistent with past mass estimates \citep[e.g.,][]{Hilton_ACT_2021}.

    \item We find that the cluster is dynamically active. We detected no line-of-sight substructures but find evidence for three subclusters in the projected distribution of member galaxies (see Fig. \ref{fig:mclust_groups} and Table \ref{tbl:mclust}). Through analysis of the PPS diagram (see Fig. \ref{fig:phase_space_mclust}), we reveal a large population of recent infallers, particularly for the southwestern subcluster, which includes the BCG. Together, the substructure and PPS analysis suggest that the cluster underwent a merger along the plane of the sky. Our findings are consistent with evidence that galaxy clusters are still accreting substructures at intermediate and low redshifts \citep[e.g.,][]{Gonzalez_assembly_2005, Carrasco_formation_2007, Dawson_merging_2015, Monteiro-Oliveira22-Hercules}.

    \item We recover the parameters of this merger using the Monte Carlo Merger Analysis Code \citep[\texttt{MCMAC},][]{Dawson_MCMAC_2013}. With an ex-post-facto prior to constrain the merger axis angle, the simulation suggests the first pericentric passage of the two most massive subclusters occurred $\sim$0.5 Gyr or $\sim$1.24 Gyr ago for an outgoing or incoming merger scenario, respectively. 
    


\end{enumerate}

We call for further investigations to better constrain the nature of J025859. For example, deeper X--ray imaging will reveal whether or not the ICM is truly dissociated from subclusters B and C as suggested by our shallow eROSITA contours (see Fig. \ref{fig:mclust_groups}). Additionally, our merger scenario analysis is significantly limited by our consideration of only two subclusters, revealing the need to study collisions between three subclusters using hydrodynamic simulations. Comparing the ICM morphology in these simulations with the eROSITA X--ray data may also help to distinguish between possible collision scenarios \citep[e.g.,][]{Monteiro-Oliveira22-Hercules, Aldas_scaling_2026}. Finally, a more complete dynamical analysis that extends beyond the $R_{200}$ radius and includes photometric redshifts could reveal the true dynamical state of J025859 by investigating the relationship between this cluster and large-scale structures, such as infalling groups, other clusters, and filaments \citep[e.g.,][]{Carrasco_dissecting_2021}.


More broadly, this study demonstrates the promise of the many disturbed galaxy cluster candidates recently revealed in the eROSITA All-Sky Survey \citepalias{Zenteno_BCG_2025} and included in eDGeS. For several of these candidates, multi-object spectroscopic data and deep optical imaging have already been collected.
A future paper will utilize a multi-wavelength suite of analyses to probe the dynamical properties, substructure, and assembly history of strongly disturbed clusters in eDGeS. Furthermore, this sample
will provide the opportunity to further test the effect of cluster mergers on galaxy evolution and further constrain the self-interaction cross-section of dark matter.




\begin{acknowledgements}

A.S.A. gratefully acknowledges the support of the Robertson Scholars Leadership Program and the hospitality of the AURA Recinto, where this work was partially done. F.A.G. acknowledges support from the ANID BASAL project FB210003, from the ANID FONDECYT Regular grant 1251493, and from the HORIZON-MSCA-2021-SE-01 Research and Innovation Programme under the Marie Sklodowska Curie grant agreement number 101086388.

This work is based on observations obtained at the international Gemini Observatory, a program of NSF's NOIRLab, which is managed by the Association of Universities for Research in Astronomy (AURA) under a cooperative agreement with the National Science Foundation on behalf of the Gemini Observatory partnership: the National Science Foundation (United States), National Research Council (Canada), Agencia Nacional de Investigación y Desarrollo (Chile), Ministerio de Ciencia, Tecnología e Innovación (Argentina), Ministério da Ciência, Tecnologia, Inovações e Comunicações (Brazil), and Korea Astronomy and Space Science Institute (Republic of Korea).

The Legacy Surveys consist of three individual and complementary projects: the Dark Energy Camera Legacy Survey (DECaLS; Proposal ID \#2014B-0404; PIs: David Schlegel and Arjun Dey), the Beijing-Arizona Sky Survey (BASS; NOAO Prop. ID \#2015A-0801; PIs: Zhou Xu and Xiaohui Fan), and the Mayall z-band Legacy Survey (MzLS; Prop. ID \#2016A-0453; PI: Arjun Dey). DECaLS, BASS and MzLS together include data obtained, respectively, at the Blanco telescope, Cerro Tololo Inter-American Observatory, NSF’s NOIRLab; the Bok telescope, Steward Observatory, University of Arizona; and the Mayall telescope, Kitt Peak National Observatory, NOIRLab. Pipeline processing and analyses of the data were supported by NOIRLab and the Lawrence Berkeley National Laboratory (LBNL). The Legacy Surveys project is honored to be permitted to conduct astronomical research on Iolkam Du’ag (Kitt Peak), a mountain with particular significance to the Tohono O’odham Nation.

NOIRLab is operated by the Association of Universities for Research in Astronomy (AURA) under a cooperative agreement with the National Science Foundation. LBNL is managed by the Regents of the University of California under contract to the U.S. Department of Energy.

This project used data obtained with the Dark Energy Camera (DECam), which was constructed by the Dark Energy Survey (DES) collaboration. Funding for the DES Projects has been provided by the U.S. Department of Energy, the U.S. National Science Foundation, the Ministry of Science and Education of Spain, the Science and Technology Facilities Council of the United Kingdom, the Higher Education Funding Council for England, the National Center for Supercomputing Applications at the University of Illinois at Urbana-Champaign, the Kavli Institute of Cosmological Physics at the University of Chicago, Center for Cosmology and Astro-Particle Physics at the Ohio State University, the Mitchell Institute for Fundamental Physics and Astronomy at Texas A\&M University, Financiadora de Estudos e Projetos, Fundacao Carlos Chagas Filho de Amparo, Financiadora de Estudos e Projetos, Fundacao Carlos Chagas Filho de Amparo a Pesquisa do Estado do Rio de Janeiro, Conselho Nacional de Desenvolvimento Cientifico e Tecnologico and the Ministerio da Ciencia, Tecnologia e Inovacao, the Deutsche Forschungsgemeinschaft and the Collaborating Institutions in the Dark Energy Survey. The Collaborating Institutions are Argonne National Laboratory, the University of California at Santa Cruz, the University of Cambridge, Centro de Investigaciones Energeticas, Medioambientales y Tecnologicas-Madrid, the University of Chicago, University College London, the DES-Brazil Consortium, the University of Edinburgh, the Eidgenossische Technische Hochschule (ETH) Zurich, Fermi National Accelerator Laboratory, the University of Illinois at Urbana-Champaign, the Institut de Ciencies de l’Espai (IEEC/CSIC), the Institut de Fisica d’Altes Energies, Lawrence Berkeley National Laboratory, the Ludwig Maximilians Universitat Munchen and the associated Excellence Cluster Universe, the University of Michigan, NSF’s NOIRLab, the University of Nottingham, the Ohio State University, the University of Pennsylvania, the University of Portsmouth, SLAC National Accelerator Laboratory, Stanford University, the University of Sussex, and Texas A\&M University.

BASS is a key project of the Telescope Access Program (TAP), which has been funded by the National Astronomical Observatories of China, the Chinese Academy of Sciences (the Strategic Priority Research Program “The Emergence of Cosmological Structures” Grant \# XDB09000000), and the Special Fund for Astronomy from the Ministry of Finance. The BASS is also supported by the External Cooperation Program of the Chinese Academy of Sciences (Grant \# 114A11KYSB20160057), and the Chinese National Natural Science Foundation (Grant \# 12120101003, \# 11433005).

The Legacy Survey team makes use of data products from the Near-Earth Object Wide-field Infrared Survey Explorer (NEOWISE), which is a project of the Jet Propulsion Laboratory/California Institute of Technology. NEOWISE is funded by the National Aeronautics and Space Administration.

The Legacy Surveys imaging of the DESI footprint is supported by the Director, Office of Science, Office of High Energy Physics of the U.S. Department of Energy under Contract No. DE-AC02-05CH1123, by the National Energy Research Scientific Computing Center, a DOE Office of Science User Facility under the same contract; and by the U.S. National Science Foundation, Division of Astronomical Sciences under Contract No. AST-0950945 to NOAO.

This work is based on data from eROSITA, the soft X--ray instrument aboard SRG, a joint Russian-German science mission supported by the Russian Space Agency (Roskosmos), in the interests of the Russian Academy of Sciences represented by its Space Research Institute (IKI), and the Deutsches Zentrum für Luft- und Raumfahrt (DLR). The SRG spacecraft was built by Lavochkin Association (NPOL) and its subcontractors, and is operated by NPOL with support from the Max Planck Institute for Extraterrestrial Physics (MPE). The development and construction of the eROSITA X--ray instrument was led by MPE, with contributions from the Dr. Karl Remeis Observatory Bamberg \& ECAP (FAU Erlangen-Nuernberg), the University of Hamburg Observatory, the Leibniz Institute for Astrophysics Potsdam (AIP), and the Institute for Astronomy and Astrophysics of the University of Tübingen, with the support of DLR and the Max Planck Society. The Argelander Institute for Astronomy of the University of Bonn and the Ludwig Maximilians Universität Munich also participated in the science preparation for eROSITA.

\facility {Gemini South: GMOS-S}
\software {\texttt{astropy v5.0} \citep{astropy}, \texttt{k-correct v5.0.0} \citep{kcorrect}, \texttt{mclust} \citep{mclust}, \texttt{MCMAC} \citep{Dawson_MCMAC_2013}
}

\end{acknowledgements}

\bibliographystyle{aasjournal}
\bibliography{Cluster}{}

@ARTICLE{Ahad26,
       author = {{Ahad}, Syeda Lammim and {Reid}, Rashaad and {Mpetha}, Charlie T. and {Taylor}, James E. and {Hildebrandt}, Hendrik and {Hudson}, Michael J. and {Chambers}, Kenneth C. and {de Boer}, Thomas and {Guerrini}, Sacha and {Guinot}, Axel and {Gwyn}, Stephen and {Kilbinger}, Martin and {Van Waerbeke}, Ludovic},
        title = "{Cluster Properties as a Function of Dynamical State in the DESI Legacy {\texttimes} UNIONS Surveys}",
      journal = {\apj},
         year = 2026,
        month = aug,
       volume = {1006},
       number = {2},
          eid = {243},
        pages = {243},
          doi = {10.3847/1538-4357/ae8399},
archivePrefix = {arXiv},
       eprint = {2512.14636},
 primaryClass = {astro-ph.GA},
       adsurl = {https://ui.adsabs.harvard.edu/abs/2026ApJ..1006..243A}
}

@ARTICLE{Randall2008,
       author = {{Randall}, Scott W. and {Markevitch}, Maxim and {Clowe}, Douglas and {Gonzalez}, Anthony H. and {Brada{\v{c}}}, Marusa},
        title = "{Constraints on the Self-Interaction Cross Section of Dark Matter from Numerical Simulations of the Merging Galaxy Cluster 1E 0657-56}",
      journal = {\apj},
         year = 2008,
        month = jun,
       volume = {679},
       number = {2},
        pages = {1173-1180},
          doi = {10.1086/587859},
archivePrefix = {arXiv},
       eprint = {0704.0261},
 primaryClass = {astro-ph},
       adsurl = {https://ui.adsabs.harvard.edu/abs/2008ApJ...679.1173R}
}

@ARTICLE{Monna2017,
       author = {{Monna}, A. and {Seitz}, S. and {Geller}, M.~J. and {Zitrin}, A. and {Mercurio}, A. and {Suyu}, S.~H. and {Postman}, M. and {Fabricant}, D.~G. and {Hwang}, H.~S. and {Koekemoer}, A.},
        title = "{Separating galaxies from the cluster dark matter halo in Abell 611}",
      journal = {\mnras},
         year = 2017,
        month = mar,
       volume = {465},
       number = {4},
        pages = {4589-4601},
          doi = {10.1093/mnras/stw3048},
archivePrefix = {arXiv},
       eprint = {1602.08491},
 primaryClass = {astro-ph.GA},
       adsurl = {https://ui.adsabs.harvard.edu/abs/2017MNRAS.465.4589M}
}

@ARTICLE{Verdugo2020,
       author = {{Verdugo}, Tom{\'a}s and {Carrasco}, Eleazar R. and {Fo{\"e}x}, Gael and
         {Motta}, Ver{\'o}nica and {Gomez}, Percy L. and {Limousin}, Marceau and
         {Maga{\~n}a}, Juan and {de Diego}, Jos{\'e} A.},
        title = "{Dissecting the Strong-lensing Galaxy Cluster MS 0440.5+0204. I. The Mass Density Profile}",
      journal = {\apj},
         year = 2020,
        month = jul,
       volume = {897},
       number = {1},
          eid = {4},
        pages = {4},
          doi = {10.3847/1538-4357/ab9635},
archivePrefix = {arXiv},
       eprint = {2006.03927},
 primaryClass = {astro-ph.CO},
       adsurl = {https://ui.adsabs.harvard.edu/abs/2020ApJ...897....4V}
}

@ARTICLE{Kravtsov2006,
       author = {{Kravtsov}, Andrey V. and {Vikhlinin}, Alexey and {Nagai}, Daisuke},
        title = "{A New Robust Low-Scatter X-Ray Mass Indicator for Clusters of Galaxies}",
      journal = {\apj},
         year = 2006,
        month = oct,
       volume = {650},
       number = {1},
        pages = {128-136},
          doi = {10.1086/506319},
archivePrefix = {arXiv},
       eprint = {astro-ph/0603205},
 primaryClass = {astro-ph},
       adsurl = {https://ui.adsabs.harvard.edu/abs/2006ApJ...650..128K}
}

@ARTICLE{Vikhlinin2009,
       author = {{Vikhlinin}, A. and {Kravtsov}, A.~V. and {Burenin}, R.~A. and {Ebeling}, H. and {Forman}, W.~R. and {Hornstrup}, A. and {Jones}, C. and {Murray}, S.~S. and {Nagai}, D. and {Quintana}, H. and {Voevodkin}, A.},
        title = "{Chandra Cluster Cosmology Project III: Cosmological Parameter Constraints}",
      journal = {\apj},
         year = 2009,
        month = feb,
       volume = {692},
       number = {2},
        pages = {1060-1074},
          doi = {10.1088/0004-637X/692/2/1060},
archivePrefix = {arXiv},
       eprint = {0812.2720},
 primaryClass = {astro-ph},
       adsurl = {https://ui.adsabs.harvard.edu/abs/2009ApJ...692.1060V}
}

@ARTICLE{Watson2026,
       author = {{Watson}, Courtney B. and {Blanton}, Elizabeth L. and {Randall}, Scott W. and {Clarke}, Tracy E. and {ZuHone}, John A.},
        title = "{Deep Chandra X-Ray Observations of A2029: The Merger History of a Relaxed, Strong Cool Core Cluster}",
      journal = {\apj},
         year = 2026,
        month = jan,
       volume = {996},
       number = {1},
          eid = {106},
        pages = {106},
          doi = {10.3847/1538-4357/ae2026},
archivePrefix = {arXiv},
       eprint = {2511.00250},
 primaryClass = {astro-ph.HE},
       adsurl = {https://ui.adsabs.harvard.edu/abs/2026ApJ...996..106W}
}

@ARTICLE{Doubrawa2023,
       author = {{Doubrawa}, Lia and {Cypriano}, Eduardo S. and {Finoguenov}, Alexis and {Lopes}, Paulo A.~A. and {Maturi}, Matteo and {Gonzalez}, Anthony H. and {Dupke}, Renato},
        title = "{Galaxy cluster optical mass proxies from probabilistic memberships}",
      journal = {\mnras},
         year = 2023,
        month = dec,
       volume = {526},
       number = {3},
        pages = {4285-4295},
          doi = {10.1093/mnras/stad3024},
archivePrefix = {arXiv},
       eprint = {2312.10564},
 primaryClass = {astro-ph.CO},
       adsurl = {https://ui.adsabs.harvard.edu/abs/2023MNRAS.526.4285D}
}

@ARTICLE{Carrasco2010,
       author = {{Carrasco}, E.~R. and {Gomez}, P.~L. and {Verdugo}, T. and {Lee}, H. and {Diaz}, R. and {Bergmann}, M. and {Turner}, J.~E.~H. and {Miller}, B.~W. and {West}, M.~J.},
        title = "{Strong Gravitational Lensing by the Super-massive cD Galaxy in Abell 3827}",
      journal = {\apjl},
         year = 2010,
        month = jun,
       volume = {715},
       number = {2},
        pages = {L160-L164},
          doi = {10.1088/2041-8205/715/2/L160},
archivePrefix = {arXiv},
       eprint = {1004.5410},
 primaryClass = {astro-ph.CO},
       adsurl = {https://ui.adsabs.harvard.edu/abs/2010ApJ...715L.160C}
}

@ARTICLE{Harvey2013,
       author = {{Harvey}, David and {Massey}, Richard and {Kitching}, Thomas and {Taylor}, Andy and {Jullo}, Eric and {Kneib}, Jean-Paul and {Tittley}, Eric and {Marshall}, Philip J.},
        title = "{Dark matter astrometry: accuracy of subhalo positions for the measurement of self-interaction cross-sections}",
      journal = {\mnras},
         year = 2013,
        month = aug,
       volume = {433},
       number = {2},
        pages = {1517-1528},
          doi = {10.1093/mnras/stt819},
archivePrefix = {arXiv},
       eprint = {1305.2117},
 primaryClass = {astro-ph.CO},
       adsurl = {https://ui.adsabs.harvard.edu/abs/2013MNRAS.433.1517H}
}

@ARTICLE{Pratt2019,
       author = {{Pratt}, G.~W. and {Arnaud}, M. and {Biviano}, A. and {Eckert}, D. and {Ettori}, S. and {Nagai}, D. and {Okabe}, N. and {Reiprich}, T.~H.},
        title = "{The Galaxy Cluster Mass Scale and Its Impact on Cosmological Constraints from the Cluster Population}",
      journal = {\ssr},
         year = 2019,
        month = feb,
       volume = {215},
       number = {2},
          eid = {25},
        pages = {25},
          doi = {10.1007/s11214-019-0591-0},
archivePrefix = {arXiv},
       eprint = {1902.10837},
 primaryClass = {astro-ph.CO},
       adsurl = {https://ui.adsabs.harvard.edu/abs/2019SSRv..215...25P}
}

@ARTICLE{Fischer2023,
       author = {{Fischer}, Moritz S. and {Durke}, Nils-Henrik and {Hollingshausen}, Katharina and {Hammer}, Claudius and {Br{\"u}ggen}, Marcus and {Dolag}, Klaus},
        title = "{The role of baryons in self-interacting dark matter mergers}",
      journal = {\mnras},
         year = 2023,
        month = aug,
       volume = {523},
       number = {4},
        pages = {5915-5933},
          doi = {10.1093/mnras/stad1786},
archivePrefix = {arXiv},
       eprint = {2302.07882},
 primaryClass = {astro-ph.CO},
       adsurl = {https://ui.adsabs.harvard.edu/abs/2023MNRAS.523.5915F}
}

@ARTICLE{Harvey2015,
       author = {{Harvey}, David and {Massey}, Richard and {Kitching}, Thomas and {Taylor}, Andy and {Tittley}, Eric},
        title = "{The nongravitational interactions of dark matter in colliding galaxy clusters}",
      journal = {Science},
         year = 2015,
        month = mar,
       volume = {347},
       number = {6229},
        pages = {1462-1465},
          doi = {10.1126/science.1261381},
archivePrefix = {arXiv},
       eprint = {1503.07675},
 primaryClass = {astro-ph.CO},
       adsurl = {https://ui.adsabs.harvard.edu/abs/2015Sci...347.1462H}
}

@ARTICLE{Markevitch2004,
       author = {{Markevitch}, M. and {Gonzalez}, A.~H. and {Clowe}, D. and {Vikhlinin}, A. and {Forman}, W. and {Jones}, C. and {Murray}, S. and {Tucker}, W.},
        title = "{Direct Constraints on the Dark Matter Self-Interaction Cross Section from the Merging Galaxy Cluster 1E 0657-56}",
      journal = {\apj},
         year = 2004,
        month = may,
       volume = {606},
       number = {2},
        pages = {819-824},
          doi = {10.1086/383178},
archivePrefix = {arXiv},
       eprint = {astro-ph/0309303},
 primaryClass = {astro-ph},
       adsurl = {https://ui.adsabs.harvard.edu/abs/2004ApJ...606..819M}
}

@ARTICLE{Owers2014,
       author = {{Owers}, Matt S. and {Nulsen}, Paul E.~J. and {Couch}, Warrick J. and {Ma}, Cheng-Jiun and {David}, Laurence P. and {Forman}, William R. and {Hopkins}, Andrew M. and {Jones}, Christine and {van Weeren}, Reinout J.},
        title = "{A Merger Shock in A2034}",
      journal = {\apj},
         year = 2014,
        month = jan,
       volume = {780},
       number = {2},
          eid = {163},
        pages = {163},
          doi = {10.1088/0004-637X/780/2/163},
archivePrefix = {arXiv},
       eprint = {1311.5561},
 primaryClass = {astro-ph.CO},
       adsurl = {https://ui.adsabs.harvard.edu/abs/2014ApJ...780..163O}
}

@ARTICLE{KravtsovBorgani2012,
       author = {{Kravtsov}, Andrey V. and {Borgani}, Stefano},
        title = "{Formation of Galaxy Clusters}",
      journal = {\araa},
         year = 2012,
        month = sep,
       volume = {50},
        pages = {353-409},
          doi = {10.1146/annurev-astro-081811-125502},
archivePrefix = {arXiv},
       eprint = {1205.5556},
 primaryClass = {astro-ph.CO},
       adsurl = {https://ui.adsabs.harvard.edu/abs/2012ARA&A..50..353K}
}

@ARTICLE{Hook_GMOS_2004,
       author = {{Hook}, I.~M. and {J{\o}rgensen}, Inger and {Allington-Smith}, J.~R. and {Davies}, R.~L. and {Metcalfe}, N. and {Murowinski}, R.~G. and {Crampton}, D.},
        title = "{The Gemini-North Multi-Object Spectrograph: Performance in Imaging, Long-Slit, and Multi-Object Spectroscopic Modes}",
      journal = {\pasp},
         year = 2004,
        month = may,
       volume = {116},
       number = {819},
        pages = {425-440},
          doi = {10.1086/383624},
       adsurl = {https://ui.adsabs.harvard.edu/abs/2004PASP..116..425H}
}

@ARTICLE{Prochaska_Pypeit_2020,
       author = {{Prochaska}, J. and {Hennawi}, Joseph and {Westfall}, Kyle and {Cooke}, Ryan and {Wang}, Feige and {Hsyu}, Tiffany and {Davies}, Frederick and {Farina}, Emanuele and {Pelliccia}, Debora},
        title = "{PypeIt: The Python Spectroscopic Data Reduction Pipeline}",
      journal = {The Journal of Open Source Software},
         year = 2020,
        month = dec,
       volume = {5},
       number = {56},
          eid = {2308},
        pages = {2308},
          doi = {10.21105/joss.02308},
archivePrefix = {arXiv},
       eprint = {2005.06505},
 primaryClass = {astro-ph.IM},
       adsurl = {https://ui.adsabs.harvard.edu/abs/2020JOSS....5.2308P}
}

@ARTICLE{Zenteno_BCG_2025,
       author = {{Zenteno}, A. and {Kluge}, M. and {Kharkrang}, R. and {Hernandez-Lang}, D. and {Damke}, G. and {Saro}, A. and {Monteiro-Oliveira}, R. and {Carrasco}, E.~R. and {Salvato}, M. and {Comparat}, J. and {Fabricius}, M. and {Snigula}, J. and {Arevalo}, P. and {Cuevas}, H. and {Nilo Castellon}, J.~L. and {Ramirez}, A. and {V{\'e}liz Astudillo}, S. and {Landriau}, M. and {Myers}, A.~D. and {Schlafly}, E. and {Valdes}, F. and {Weaver}, B.~A. and {Mohr}, J.~J. and {Grandis}, S. and {Klein}, M. and {Liu}, A. and {Bulbul}, E. and {Zhang}, X. and {Sanders}, J.~S. and {Bahar}, Y.~E. and {Ghirardini}, V. and {Ramos}, M.~E. and {Balzer}, F.},
        title = "{The dynamical state of eROSITA clusters and its impact on the brightest cluster galaxy luminosity}",
      journal = {\aap},
         year = 2025,
        month = jun,
       volume = {698},
          eid = {A171},
        pages = {A171},
          doi = {10.1051/0004-6361/202452440},
archivePrefix = {arXiv},
       eprint = {2503.21066},
 primaryClass = {astro-ph.CO},
       adsurl = {https://ui.adsabs.harvard.edu/abs/2025A&A...698A.171Z}
}

@ARTICLE{Kluge_optical_2024,
       author = {{Kluge}, M. and {Comparat}, J. and {Liu}, A. and {Balzer}, F. and {Bulbul}, E. and {Ider Chitham}, J. and {Ghirardini}, V. and {Garrel}, C. and {Bahar}, Y.~E. and {Artis}, E. and {Bender}, R. and {Clerc}, N. and {Dwelly}, T. and {Fabricius}, M.~H. and {Grandis}, S. and {Hern{\'a}ndez-Lang}, D. and {Hill}, G.~J. and {Joshi}, J. and {Lamer}, G. and {Merloni}, A. and {Nandra}, K. and {Pacaud}, F. and {Predehl}, P. and {Ramos-Ceja}, M.~E. and {Reiprich}, T.~H. and {Salvato}, M. and {Sanders}, J.~S. and {Schrabback}, T. and {Seppi}, R. and {Zelmer}, S. and {Zenteno}, A. and {Zhang}, X.},
        title = "{The SRG/eROSITA All-Sky Survey. Optical identification and properties of galaxy clusters and groups in the western galactic hemisphere}",
      journal = {\aap},
         year = 2024,
        month = aug,
       volume = {688},
          eid = {A210},
        pages = {A210},
          doi = {10.1051/0004-6361/202349031},
archivePrefix = {arXiv},
       eprint = {2402.08453},
 primaryClass = {astro-ph.CO},
       adsurl = {https://ui.adsabs.harvard.edu/abs/2024A&A...688A.210K}
}

@ARTICLE{Bulbul_clusters_2024,
       author = {{Bulbul}, E. and {Liu}, A. and {Kluge}, M. and {Zhang}, X. and {Sanders}, J.~S. and {Bahar}, Y.~E. and {Ghirardini}, V. and {Artis}, E. and {Seppi}, R. and {Garrel}, C. and {Ramos-Ceja}, M.~E. and {Comparat}, J. and {Balzer}, F. and {B{\"o}ckmann}, K. and {Br{\"u}ggen}, M. and {Clerc}, N. and {Dennerl}, K. and {Dolag}, K. and {Freyberg}, M. and {Grandis}, S. and {Gruen}, D. and {Kleinebreil}, F. and {Krippendorf}, S. and {Lamer}, G. and {Merloni}, A. and {Migkas}, K. and {Nandra}, K. and {Pacaud}, F. and {Predehl}, P. and {Reiprich}, T.~H. and {Schrabback}, T. and {Veronica}, A. and {Weller}, J. and {Zelmer}, S.},
        title = "{The SRG/eROSITA All-Sky Survey. The first catalog of galaxy clusters and groups in the Western Galactic Hemisphere}",
      journal = {\aap},
         year = 2024,
        month = may,
       volume = {685},
          eid = {A106},
        pages = {A106},
          doi = {10.1051/0004-6361/202348264},
archivePrefix = {arXiv},
       eprint = {2402.08452},
 primaryClass = {astro-ph.CO},
       adsurl = {https://ui.adsabs.harvard.edu/abs/2024A&A...685A.106B}
}

@ARTICLE{DES21A,
       author = {{Abbott}, T.~M.~C. and {Adam{\'o}w}, M. and {Aguena}, M. and {Allam}, S. and {Amon}, A. and {Annis}, J. and {Avila}, S. and {Bacon}, D. and {Banerji}, M. and {Bechtol}, K. and {Becker}, M.~R. and {Bernstein}, G.~M. and {Bertin}, E. and {Bhargava}, S. and {Bridle}, S.~L. and {Brooks}, D. and {Burke}, D.~L. and {Carnero Rosell}, A. and {Carrasco Kind}, M. and {Carretero}, J. and {Castander}, F.~J. and {Cawthon}, R. and {Chang}, C. and {Choi}, A. and {Conselice}, C. and {Costanzi}, M. and {Crocce}, M. and {da Costa}, L.~N. and {Davis}, T.~M. and {De Vicente}, J. and {DeRose}, J. and {Desai}, S. and {Diehl}, H.~T. and {Dietrich}, J.~P. and {Drlica-Wagner}, A. and {Eckert}, K. and {Elvin-Poole}, J. and {Everett}, S. and {Evrard}, A.~E. and {Ferrero}, I. and {Fert{\'e}}, A. and {Flaugher}, B. and {Fosalba}, P. and {Friedel}, D. and {Frieman}, J. and {Garc{\'\i}a-Bellido}, J. and {Gaztanaga}, E. and {Gelman}, L. and {Gerdes}, D.~W. and {Giannantonio}, T. and {Gill}, M.~S.~S. and {Gruen}, D. and {Gruendl}, R.~A. and {Gschwend}, J. and {Gutierrez}, G. and {Hartley}, W.~G. and {Hinton}, S.~R. and {Hollowood}, D.~L. and {Honscheid}, K. and {Huterer}, D. and {James}, D.~J. and {Jeltema}, T. and {Johnson}, M.~D. and {Kent}, S. and {Kron}, R. and {Kuehn}, K. and {Kuropatkin}, N. and {Lahav}, O. and {Li}, T.~S. and {Lidman}, C. and {Lin}, H. and {MacCrann}, N. and {Maia}, M.~A.~G. and {Manning}, T.~A. and {Maloney}, J.~D. and {March}, M. and {Marshall}, J.~L. and {Martini}, P. and {Melchior}, P. and {Menanteau}, F. and {Miquel}, R. and {Morgan}, R. and {Myles}, J. and {Neilsen}, E. and {Ogando}, R.~L.~C. and {Palmese}, A. and {Paz-Chinch{\'o}n}, F. and {Petravick}, D. and {Pieres}, A. and {Plazas}, A.~A. and {Pond}, C. and {Rodriguez-Monroy}, M. and {Romer}, A.~K. and {Roodman}, A. and {Rykoff}, E.~S. and {Sako}, M. and {Sanchez}, E. and {Santiago}, B. and {Scarpine}, V. and {Serrano}, S. and {Sevilla-Noarbe}, I. and {Smith}, J. Allyn and {Smith}, M. and {Soares-Santos}, M. and {Suchyta}, E. and {Swanson}, M.~E.~C. and {Tarle}, G. and {Thomas}, D. and {To}, C. and {Tremblay}, P.~E. and {Troxel}, M.~A. and {Tucker}, D.~L. and {Turner}, D.~J. and {Varga}, T.~N. and {Walker}, A.~R. and {Wechsler}, R.~H. and {Weller}, J. and {Wester}, W. and {Wilkinson}, R.~D. and {Yanny}, B. and {Zhang}, Y. and {Nikutta}, R. and {Fitzpatrick}, M. and {Jacques}, A. and {Scott}, A. and {Olsen}, K. and {Huang}, L. and {Herrera}, D. and {Juneau}, S. and {Nidever}, D. and {Weaver}, B.~A. and {Adean}, C. and {Correia}, V. and {de Freitas}, M. and {Freitas}, F.~N. and {Singulani}, C. and {Vila-Verde}, G. and {Linea Science Server}},
        title = "{The Dark Energy Survey Data Release 2}",
      journal = {\apjs},
         year = 2021,
        month = aug,
       volume = {255},
       number = {2},
          eid = {20},
        pages = {20},
          doi = {10.3847/1538-4365/ac00b3},
archivePrefix = {arXiv},
       eprint = {2101.05765},
 primaryClass = {astro-ph.IM},
       adsurl = {https://ui.adsabs.harvard.edu/abs/2021ApJS..255...20A}
}

@ARTICLE{Sanders_morphologies_2025,
       author = {{Sanders}, J.~S. and {Bahar}, Y.~E. and {Bulbul}, E. and {Ghirardini}, V. and {Liu}, A. and {Clerc}, N. and {Ramos-Ceja}, M.~E. and {Reiprich}, T.~H. and {Balzer}, F. and {Comparat}, J. and {Kluge}, M. and {Pacaud}, F. and {Zhang}, X.},
        title = "{The SRG/eROSITA all-sky survey: The morphologies of clusters of galaxies: I. A catalogue of morphological parameters}",
      journal = {\aap},
         year = 2025,
        month = mar,
       volume = {695},
          eid = {A160},
        pages = {A160},
          doi = {10.1051/0004-6361/202452618},
archivePrefix = {arXiv},
       eprint = {2502.02239},
 primaryClass = {astro-ph.CO},
       adsurl = {https://ui.adsabs.harvard.edu/abs/2025A&A...695A.160S}
}

@ARTICLE{Bocquet_cosmology_2015,
       author = {{Bocquet}, S. and {Saro}, A. and {Mohr}, J.~J. and {Aird}, K.~A. and {Ashby}, M.~L.~N. and {Bautz}, M. and {Bayliss}, M. and {Bazin}, G. and {Benson}, B.~A. and {Bleem}, L.~E. and {Brodwin}, M. and {Carlstrom}, J.~E. and {Chang}, C.~L. and {Chiu}, I. and {Cho}, H.~M. and {Clocchiatti}, A. and {Crawford}, T.~M. and {Crites}, A.~T. and {Desai}, S. and {de Haan}, T. and {Dietrich}, J.~P. and {Dobbs}, M.~A. and {Foley}, R.~J. and {Forman}, W.~R. and {Gangkofner}, D. and {George}, E.~M. and {Gladders}, M.~D. and {Gonzalez}, A.~H. and {Halverson}, N.~W. and {Hennig}, C. and {Hlavacek-Larrondo}, J. and {Holder}, G.~P. and {Holzapfel}, W.~L. and {Hrubes}, J.~D. and {Jones}, C. and {Keisler}, R. and {Knox}, L. and {Lee}, A.~T. and {Leitch}, E.~M. and {Liu}, J. and {Lueker}, M. and {Luong-Van}, D. and {Marrone}, D.~P. and {McDonald}, M. and {McMahon}, J.~J. and {Meyer}, S.~S. and {Mocanu}, L. and {Murray}, S.~S. and {Padin}, S. and {Pryke}, C. and {Reichardt}, C.~L. and {Rest}, A. and {Ruel}, J. and {Ruhl}, J.~E. and {Saliwanchik}, B.~R. and {Sayre}, J.~T. and {Schaffer}, K.~K. and {Shirokoff}, E. and {Spieler}, H.~G. and {Stalder}, B. and {Stanford}, S.~A. and {Staniszewski}, Z. and {Stark}, A.~A. and {Story}, K. and {Stubbs}, C.~W. and {Vanderlinde}, K. and {Vieira}, J.~D. and {Vikhlinin}, A. and {Williamson}, R. and {Zahn}, O. and {Zenteno}, A.},
        title = "{Mass Calibration and Cosmological Analysis of the SPT-SZ Galaxy Cluster Sample Using Velocity Dispersion {\ensuremath{\sigma}}$_{ v }$ and X-Ray Y $_{X}$ Measurements}",
      journal = {\apj},
         year = 2015,
        month = feb,
       volume = {799},
       number = {2},
          eid = {214},
        pages = {214},
          doi = {10.1088/0004-637X/799/2/214},
archivePrefix = {arXiv},
       eprint = {1407.2942},
 primaryClass = {astro-ph.CO},
       adsurl = {https://ui.adsabs.harvard.edu/abs/2015ApJ...799..214B}
}

@ARTICLE{Ribero_PPS_2023,
       author = {{Ribeiro}, A.~L.~B. and {Nascimento}, R.~S. and {Morell}, D.~F. and {Lopes}, P.~A.~A. and {Dantas}, C.~C. and {Fonseca}, M.~H.~S.},
        title = "{Late growth of early-type galaxies in low-z massive clusters}",
      journal = {\mnras},
         year = 2023,
        month = may,
       volume = {521},
       number = {1},
        pages = {1221-1232},
          doi = {10.1093/mnras/stad468},
archivePrefix = {arXiv},
       eprint = {2302.04287},
 primaryClass = {astro-ph.GA},
       adsurl = {https://ui.adsabs.harvard.edu/abs/2023MNRAS.521.1221R}
}

@ARTICLE{Carrasco_dissecting_2021,
       author = {{Carrasco}, Eleazar R. and {Verdugo}, Tom{\'a}s and {Motta}, Ver{\'o}nica and {Fo{\"e}x}, Gael and {Ellingson}, E. and {Gomez}, Percy L. and {Falco}, Emilio and {Limousin}, Marceau},
        title = "{Dissecting the Strong-lensing Galaxy Cluster MS 0440.5+0204. II. New Optical Spectroscopic Observations in a Wider Area and Cluster Dynamical State}",
      journal = {\apj},
         year = 2021,
        month = sep,
       volume = {918},
       number = {2},
          eid = {61},
        pages = {61},
          doi = {10.3847/1538-4357/ac0c1b},
archivePrefix = {arXiv},
       eprint = {2106.15583},
 primaryClass = {astro-ph.GA},
       adsurl = {https://ui.adsabs.harvard.edu/abs/2021ApJ...918...61C}
}

@ARTICLE{Carlberg_mass_1997,
       author = {{Carlberg}, R.~G. and {Yee}, H.~K.~C. and {Ellingson}, E. and {Morris}, S.~L. and {Abraham}, R. and {Gravel}, P. and {Pritchet}, C.~J. and {Smecker-Hane}, T. and {Hartwick}, F.~D.~A. and {Hesser}, J.~E. and {Hutchings}, J.~B. and {Oke}, J.~B.},
        title = "{The Average Mass Profile of Galaxy Clusters}",
      journal = {\apjl},
         year = 1997,
        month = aug,
       volume = {485},
       number = {1},
        pages = {L13-L16},
          doi = {10.1086/310801},
archivePrefix = {arXiv},
       eprint = {astro-ph/9703107},
 primaryClass = {astro-ph},
       adsurl = {https://ui.adsabs.harvard.edu/abs/1997ApJ...485L..13C}
}

@ARTICLE{Munari_mass_2013,
       author = {{Munari}, E. and {Biviano}, A. and {Borgani}, S. and {Murante}, G. and {Fabjan}, D.},
        title = "{The relation between velocity dispersion and mass in simulated clusters of galaxies: dependence on the tracer and the baryonic physics}",
      journal = {\mnras},
         year = 2013,
        month = apr,
       volume = {430},
       number = {4},
        pages = {2638-2649},
          doi = {10.1093/mnras/stt049},
archivePrefix = {arXiv},
       eprint = {1301.1682},
 primaryClass = {astro-ph.CO},
       adsurl = {https://ui.adsabs.harvard.edu/abs/2013MNRAS.430.2638M}
}

@ARTICLE{Beers_location_1990,
       author = {{Beers}, Timothy C. and {Flynn}, Kevin and {Gebhardt}, Karl},
        title = "{Measures of Location and Scale for Velocities in Clusters of Galaxies---A Robust Approach}",
      journal = {\aj},
         year = 1990,
        month = jul,
       volume = {100},
        pages = {32},
          doi = {10.1086/115487},
       adsurl = {https://ui.adsabs.harvard.edu/abs/1990AJ....100...32B}
}

@ARTICLE{Gladders_method_2000,
       author = {{Gladders}, Michael D. and {Yee}, H.~K.~C.},
        title = "{A New Method For Galaxy Cluster Detection. I. The Algorithm}",
      journal = {\aj},
         year = 2000,
        month = oct,
       volume = {120},
       number = {4},
        pages = {2148-2162},
          doi = {10.1086/301557},
archivePrefix = {arXiv},
       eprint = {astro-ph/0004092},
 primaryClass = {astro-ph},
       adsurl = {https://ui.adsabs.harvard.edu/abs/2000AJ....120.2148G}
}

@ARTICLE{Gladders_sequence_1998,
       author = {{Gladders}, Michael D. and {L{\'o}pez-Cruz}, Omar and {Yee}, H.~K.~C. and {Kodama}, Tadayuki},
        title = "{The Slope of the Cluster Elliptical Red Sequence: A Probe of Cluster Evolution}",
      journal = {\apj},
         year = 1998,
        month = jul,
       volume = {501},
       number = {2},
        pages = {571-577},
          doi = {10.1086/305858},
archivePrefix = {arXiv},
       eprint = {astro-ph/9802167},
 primaryClass = {astro-ph},
       adsurl = {https://ui.adsabs.harvard.edu/abs/1998ApJ...501..571G}
}

@book{Silverman_KDE_1998,
    author =  {{Silverman}, B. W.},
    title = "{Density Estimation for Statistics and Data Analysis (1st ed.)}",
    publisher = {Routledge},
    year = 1998,
    doi = {10.1201/9781315140919}
}

@article{Scrucca_Mclust_2016,
  author = {Luca Scrucca and Michael Fop and T. Brendan Murphy and
          Adrian E. Raftery},
  title = {{mclust 5: Clustering, Classification and Density Estimation
          Using Gaussian Finite Mixture Models}},
  year = {2016},
  journal = {{The R Journal}},
  doi = {10.32614/RJ-2016-021},
  url = {https://doi.org/10.32614/RJ-2016-021},
  pages = {289--317},
  volume = {8},
  number = {1}
}

@ARTICLE{Monteiro-Oliveira_A1644_2020,
       author = {{Monteiro-Oliveira}, R. and {Doubrawa}, L. and {Machado}, R.~E.~G. and {Lima Neto}, G.~B. and {Castejon}, M. and {Cypriano}, E.~S.},
        title = "{Revising the merger scenario of the galaxy cluster Abell 1644: a new gas poor structure discovered by weak gravitational lensing}",
      journal = {\mnras},
         year = 2020,
        month = jun,
       volume = {495},
       number = {2},
        pages = {2007-2021},
          doi = {10.1093/mnras/staa1218},
archivePrefix = {arXiv},
       eprint = {2004.13662},
 primaryClass = {astro-ph.GA},
       adsurl = {https://ui.adsabs.harvard.edu/abs/2020MNRAS.495.2007M}
}

@ARTICLE{Kim_merging_2024,
       author = {{Kim}, Duho and {Sheen}, Yun-Kyeong and {Jaff{\'e}}, Yara L. and {Kelkar}, Kshitija and {Ranjan}, Adarsh and {Piraino-Cerda}, Franco and {Crossett}, Jacob P. and {Costa Louren{\c{c}}o}, Ana Carolina and {Martin}, Garreth and {Nantais}, Julie B. and {Demarco}, Ricardo and {Treister}, Ezequiel and {Yi}, Sukyoung K.},
        title = "{Distribution of Merging and Post-merger Galaxies in Nearby Galaxy Clusters}",
      journal = {\apj},
         year = 2024,
        month = may,
       volume = {966},
       number = {1},
          eid = {124},
        pages = {124},
          doi = {10.3847/1538-4357/ad32ce},
archivePrefix = {arXiv},
       eprint = {2403.06437},
 primaryClass = {astro-ph.GA},
       adsurl = {https://ui.adsabs.harvard.edu/abs/2024ApJ...966..124K}
}

@ARTICLE{Kelkar2020,
       author = {{Kelkar}, Kshitija and {Dwarakanath}, K.~S. and {Poggianti}, Bianca M. and {Moretti}, Alessia and {Monteiro-Oliveira}, Rog{\'e}rio and {Machado}, Rubens E.~G. and {Lima-Neto}, Gast{\~a}o B. and {Fritz}, Jacopo and {Vulcani}, Benedetta and {Gullieuszik}, Marco and {Bettoni}, Daniela},
        title = "{Passive spirals and shock influenced star formation in the merging cluster A3376}",
      journal = {\mnras},
         year = 2020,
        month = jul,
       volume = {496},
       number = {1},
        pages = {442-455},
          doi = {10.1093/mnras/staa1547},
archivePrefix = {arXiv},
       eprint = {2002.08610},
 primaryClass = {astro-ph.GA},
       adsurl = {https://ui.adsabs.harvard.edu/abs/2020MNRAS.496..442K}
}

@ARTICLE{PirainoCerda2024,
       author = {{Piraino-Cerda}, Franco and {Jaff{\'e}}, Yara L. and {Louren{\c{c}}o}, Ana C. and {Crossett}, Jacob P. and {Salinas}, Vicente and {Kim}, Duho and {Sheen}, Yun-Kyeong and {Kelkar}, Kshitija and {Pallero}, Diego and {Bravo-Alfaro}, Hector},
        title = "{Pre- and post-processing of cluster galaxies out to 5 {\texttimes} R$_{200}$: the extreme case of A2670}",
      journal = {\mnras},
         year = 2024,
        month = feb,
       volume = {528},
       number = {1},
        pages = {919-936},
          doi = {10.1093/mnras/stad3957},
archivePrefix = {arXiv},
       eprint = {2401.06973},
 primaryClass = {astro-ph.GA},
       adsurl = {https://ui.adsabs.harvard.edu/abs/2024MNRAS.528..919P}
}

@ARTICLE{Lourenco_bullet_2020,
       author = {{Louren{\c{c}}o}, Ana C.~C. and {Lopes}, P.~A.~A. and {Lagan{\'a}}, T.~F. and {Nascimento}, R.~S. and {Machado}, R.~E.~G. and {Moura}, M.~T. and {Jaff{\'e}}, Y.~L. and {Ribeiro}, A.~L. and {Vulcani}, B. and {Moretti}, A. and {Riguccini}, L.~A.},
        title = "{The dynamical state of Abell 2399: a bullet-like cluster}",
      journal = {\mnras},
         year = 2020,
        month = oct,
       volume = {498},
       number = {1},
        pages = {835-849},
          doi = {10.1093/mnras/staa2464},
archivePrefix = {arXiv},
       eprint = {2009.01400},
 primaryClass = {astro-ph.GA},
       adsurl = {https://ui.adsabs.harvard.edu/abs/2020MNRAS.498..835L}
}

@ARTICLE{Biernacki_ICL_2000,
  author={Biernacki, C. and Celeux, G. and Govaert, G.},
  journal={IEEE Transactions on Pattern Analysis and Machine Intelligence}, 
  title={Assessing a mixture model for clustering with the integrated completed likelihood}, 
  year={2000},
  volume={22},
  number={7},
  pages={719-725},
  doi={10.1109/34.865189}}

@ARTICLE{Wainer_regression_1976,
       author = {{Wainer}, H. and {Thissen}, D.},
        title = "{Three steps towards robust regression}",
      journal = {Psychometrika},
         year = 1976,
       volume = {41},
        pages = {9-34},
          doi = {10.1007/BF02291695},
}

@ARTICLE{Ruel_SPT_2014,
       author = {{Ruel}, J. and {Bazin}, G. and {Bayliss}, M. and {Brodwin}, M. and {Foley}, R.~J. and {Stalder}, B. and {Aird}, K.~A. and {Armstrong}, R. and {Ashby}, M.~L.~N. and {Bautz}, M. and {Benson}, B.~A. and {Bleem}, L.~E. and {Bocquet}, S. and {Carlstrom}, J.~E. and {Chang}, C.~L. and {Chapman}, S.~C. and {Cho}, H.~M. and {Clocchiatti}, A. and {Crawford}, T.~M. and {Crites}, A.~T. and {de Haan}, T. and {Desai}, S. and {Dobbs}, M.~A. and {Dudley}, J.~P. and {Forman}, W.~R. and {George}, E.~M. and {Gladders}, M.~D. and {Gonzalez}, A.~H. and {Halverson}, N.~W. and {Harrington}, N.~L. and {High}, F.~W. and {Holder}, G.~P. and {Holzapfel}, W.~L. and {Hrubes}, J.~D. and {Jones}, C. and {Joy}, M. and {Keisler}, R. and {Knox}, L. and {Lee}, A.~T. and {Leitch}, E.~M. and {Liu}, J. and {Lueker}, M. and {Luong-Van}, D. and {Mantz}, A. and {Marrone}, D.~P. and {McDonald}, M. and {McMahon}, J.~J. and {Mehl}, J. and {Meyer}, S.~S. and {Mocanu}, L. and {Mohr}, J.~J. and {Montroy}, T.~E. and {Murray}, S.~S. and {Natoli}, T. and {Nurgaliev}, D. and {Padin}, S. and {Plagge}, T. and {Pryke}, C. and {Reichardt}, C.~L. and {Rest}, A. and {Ruhl}, J.~E. and {Saliwanchik}, B.~R. and {Saro}, A. and {Sayre}, J.~T. and {Schaffer}, K.~K. and {Shaw}, L. and {Shirokoff}, E. and {Song}, J. and {{\v{S}}uhada}, R. and {Spieler}, H.~G. and {Stanford}, S.~A. and {Staniszewski}, Z. and {Starsk}, A.~A. and {Story}, K. and {Stubbs}, C.~W. and {van Engelen}, A. and {Vanderlinde}, K. and {Vieira}, J.~D. and {Vikhlinin}, A. and {Williamson}, R. and {Zahn}, O. and {Zenteno}, A.},
        title = "{Optical Spectroscopy and Velocity Dispersions of Galaxy Clusters from the SPT-SZ Survey}",
      journal = {\apj},
         year = 2014,
        month = sep,
       volume = {792},
       number = {1},
          eid = {45},
        pages = {45},
          doi = {10.1088/0004-637X/792/1/45},
archivePrefix = {arXiv},
       eprint = {1311.4953},
 primaryClass = {astro-ph.CO},
       adsurl = {https://ui.adsabs.harvard.edu/abs/2014ApJ...792...45R}
}

@ARTICLE{Noble_PhaseSpace_2016,
       author = {{Noble}, A.~G. and {Webb}, T.~M.~A. and {Yee}, H.~K.~C. and {Muzzin}, A. and {Wilson}, G. and {van der Burg}, R.~F.~J. and {Balogh}, M.~L. and {Shupe}, D.~L.},
        title = "{The Phase Space of z{\ensuremath{\sim}}1.2 SpARCS Clusters: Using Herschel to Probe Dust Temperature as a Function of Environment and Accretion History}",
      journal = {\apj},
         year = 2016,
        month = jan,
       volume = {816},
       number = {2},
          eid = {48},
        pages = {48},
          doi = {10.3847/0004-637X/816/2/48},
archivePrefix = {arXiv},
       eprint = {1511.00584},
 primaryClass = {astro-ph.GA},
       adsurl = {https://ui.adsabs.harvard.edu/abs/2016ApJ...816...48N}
}

@ARTICLE{Gill_evolution_2005,
       author = {{Gill}, Stuart P.~D. and {Knebe}, Alexander and {Gibson}, Brad K.},
        title = "{The evolution of substructure - III. The outskirts of clusters}",
      journal = {\mnras},
         year = 2005,
        month = feb,
       volume = {356},
       number = {4},
        pages = {1327-1332},
          doi = {10.1111/j.1365-2966.2004.08562.x},
archivePrefix = {arXiv},
       eprint = {astro-ph/0404427},
 primaryClass = {astro-ph},
       adsurl = {https://ui.adsabs.harvard.edu/abs/2005MNRAS.356.1327G}
}

@ARTICLE{Mahajan_backsplash_2011,
       author = {{Mahajan}, Smriti and {Mamon}, Gary A. and {Raychaudhury}, Somak},
        title = "{The velocity modulation of galaxy properties in and near clusters: quantifying the decrease in star formation in backsplash galaxies}",
      journal = {\mnras},
         year = 2011,
        month = oct,
       volume = {416},
       number = {4},
        pages = {2882-2902},
          doi = {10.1111/j.1365-2966.2011.19236.x},
archivePrefix = {arXiv},
       eprint = {1106.3062},
 primaryClass = {astro-ph.CO},
       adsurl = {https://ui.adsabs.harvard.edu/abs/2011MNRAS.416.2882M}
}

@ARTICLE{Haines_LoCuSS_2012,
       author = {{Haines}, C.~P. and {Pereira}, M.~J. and {Sanderson}, A.~J.~R. and {Smith}, G.~P. and {Egami}, E. and {Babul}, A. and {Edge}, A.~C. and {Finoguenov}, A. and {Moran}, S.~M. and {Okabe}, N.},
        title = "{LoCuSS: A Dynamical Analysis of X-Ray Active Galactic Nuclei in Local Clusters}",
      journal = {\apj},
         year = 2012,
        month = aug,
       volume = {754},
       number = {2},
          eid = {97},
        pages = {97},
          doi = {10.1088/0004-637X/754/2/97},
archivePrefix = {arXiv},
       eprint = {1205.6818},
 primaryClass = {astro-ph.CO},
       adsurl = {https://ui.adsabs.harvard.edu/abs/2012ApJ...754...97H}
}

@ARTICLE{Noble_Kinematic_2013,
       author = {{Noble}, A.~G. and {Webb}, T.~M.~A. and {Muzzin}, A. and {Wilson}, G. and {Yee}, H.~K.~C. and {van der Burg}, R.~F.~J.},
        title = "{A Kinematic Approach to Assessing Environmental Effects: Star-forming Galaxies in a z \raisebox{-0.5ex}\textasciitilde 0.9 SpARCS Cluster Using Spitzer 24 {\ensuremath{\mu}}m Observations}",
      journal = {\apj},
         year = 2013,
        month = may,
       volume = {768},
       number = {2},
          eid = {118},
        pages = {118},
          doi = {10.1088/0004-637X/768/2/118},
archivePrefix = {arXiv},
       eprint = {1303.4997},
 primaryClass = {astro-ph.CO},
       adsurl = {https://ui.adsabs.harvard.edu/abs/2013ApJ...768..118N}
}

@ARTICLE{Rhee_PhaseSpace_2017,
       author = {{Rhee}, Jinsu and {Smith}, Rory and {Choi}, Hoseung and {Yi}, Sukyoung K. and {Jaff{\'e}}, Yara and {Candlish}, Graeme and {S{\'a}nchez-J{\'a}nssen}, Ruben},
        title = "{Phase-space Analysis in the Group and Cluster Environment: Time Since Infall and Tidal Mass Loss}",
      journal = {\apj},
         year = 2017,
        month = jul,
       volume = {843},
       number = {2},
          eid = {128},
        pages = {128},
          doi = {10.3847/1538-4357/aa6d6c},
archivePrefix = {arXiv},
       eprint = {1704.04243},
 primaryClass = {astro-ph.GA},
       adsurl = {https://ui.adsabs.harvard.edu/abs/2017ApJ...843..128R}
}

@ARTICLE{Abell_catalog_1989,
       author = {{Abell}, George O. and {Corwin}, Jr., Harold G. and {Olowin}, Ronald P.},
        title = "{A Catalog of Rich Clusters of Galaxies}",
      journal = {\apjs},
         year = 1989,
        month = may,
       volume = {70},
        pages = {1},
          doi = {10.1086/191333},
       adsurl = {https://ui.adsabs.harvard.edu/abs/1989ApJS...70....1A}
}

@ARTICLE{Hilton_ACT_2021,
       author = {{Hilton}, M. and {Sif{\'o}n}, C. and {Naess}, S. and {Madhavacheril}, M. and {Oguri}, M. and {Rozo}, E. and {Rykoff}, E. and {Abbott}, T.~M.~C. and {Adhikari}, S. and {Aguena}, M. and {Aiola}, S. and {Allam}, S. and {Amodeo}, S. and {Amon}, A. and {Annis}, J. and {Ansarinejad}, B. and {Aros-Bunster}, C. and {Austermann}, J.~E. and {Avila}, S. and {Bacon}, D. and {Battaglia}, N. and {Beall}, J.~A. and {Becker}, D.~T. and {Bernstein}, G.~M. and {Bertin}, E. and {Bhandarkar}, T. and {Bhargava}, S. and {Bond}, J.~R. and {Brooks}, D. and {Burke}, D.~L. and {Calabrese}, E. and {Carrasco Kind}, M. and {Carretero}, J. and {Choi}, S.~K. and {Choi}, A. and {Conselice}, C. and {da Costa}, L.~N. and {Costanzi}, M. and {Crichton}, D. and {Crowley}, K.~T. and {D{\"u}nner}, R. and {Denison}, E.~V. and {Devlin}, M.~J. and {Dicker}, S.~R. and {Diehl}, H.~T. and {Dietrich}, J.~P. and {Doel}, P. and {Duff}, S.~M. and {Duivenvoorden}, A.~J. and {Dunkley}, J. and {Everett}, S. and {Ferraro}, S. and {Ferrero}, I. and {Fert{\'e}}, A. and {Flaugher}, B. and {Frieman}, J. and {Gallardo}, P.~A. and {Garc{\'\i}a-Bellido}, J. and {Gaztanaga}, E. and {Gerdes}, D.~W. and {Giles}, P. and {Golec}, J.~E. and {Gralla}, M.~B. and {Grandis}, S. and {Gruen}, D. and {Gruendl}, R.~A. and {Gschwend}, J. and {Gutierrez}, G. and {Han}, D. and {Hartley}, W.~G. and {Hasselfield}, M. and {Hill}, J.~C. and {Hilton}, G.~C. and {Hincks}, A.~D. and {Hinton}, S.~R. and {Ho}, S. -P.~P. and {Honscheid}, K. and {Hoyle}, B. and {Hubmayr}, J. and {Huffenberger}, K.~M. and {Hughes}, J.~P. and {Jaelani}, A.~T. and {Jain}, B. and {James}, D.~J. and {Jeltema}, T. and {Kent}, S. and {Knowles}, K. and {Koopman}, B.~J. and {Kuehn}, K. and {Lahav}, O. and {Lima}, M. and {Lin}, Y. -T. and {Lokken}, M. and {Loubser}, S.~I. and {MacCrann}, N. and {Maia}, M.~A.~G. and {Marriage}, T.~A. and {Martin}, J. and {McMahon}, J. and {Melchior}, P. and {Menanteau}, F. and {Miquel}, R. and {Miyatake}, H. and {Moodley}, K. and {Morgan}, R. and {Mroczkowski}, T. and {Nati}, F. and {Newburgh}, L.~B. and {Niemack}, M.~D. and {Nishizawa}, A.~J. and {Ogando}, R.~L.~C. and {Orlowski-Scherer}, J. and {Page}, L.~A. and {Palmese}, A. and {Partridge}, B. and {Paz-Chinch{\'o}n}, F. and {Phakathi}, P. and {Plazas}, A.~A. and {Robertson}, N.~C. and {Romer}, A.~K. and {Carnero Rosell}, A. and {Salatino}, M. and {Sanchez}, E. and {Schaan}, E. and {Schillaci}, A. and {Sehgal}, N. and {Serrano}, S. and {Shin}, T. and {Simon}, S.~M. and {Smith}, M. and {Soares-Santos}, M. and {Spergel}, D.~N. and {Staggs}, S.~T. and {Storer}, E.~R. and {Suchyta}, E. and {Swanson}, M.~E.~C. and {Tarle}, G. and {Thomas}, D. and {To}, C. and {Trac}, H. and {Ullom}, J.~N. and {Vale}, L.~R. and {Van Lanen}, J. and {Vavagiakis}, E.~M. and {De Vicente}, J. and {Wilkinson}, R.~D. and {Wollack}, E.~J. and {Xu}, Z. and {Zhang}, Y.},
        title = "{The Atacama Cosmology Telescope: A Catalog of >4000 Sunyaev-Zel{\textquoteright}dovich Galaxy Clusters}",
      journal = {\apjs},
         year = 2021,
        month = mar,
       volume = {253},
       number = {1},
          eid = {3},
        pages = {3},
          doi = {10.3847/1538-4365/abd023},
archivePrefix = {arXiv},
       eprint = {2009.11043},
 primaryClass = {astro-ph.CO},
       adsurl = {https://ui.adsabs.harvard.edu/abs/2021ApJS..253....3H}
}

@ARTICLE{Song_A2107_2018,
       author = {{Song}, Hyunmi and {Hwang}, Ho Seong and {Park}, Changbom and {Smith}, Rory and {Einasto}, Maret},
        title = "{A Redshift Survey of the Nearby Galaxy Cluster A2107: Global Rotation of the Cluster and Its Connection to Large-scale Structures in the Universe}",
      journal = {\apj},
         year = 2018,
        month = dec,
       volume = {869},
       number = {2},
          eid = {124},
        pages = {124},
          doi = {10.3847/1538-4357/aaed27},
archivePrefix = {arXiv},
       eprint = {1810.11985},
 primaryClass = {astro-ph.GA},
       adsurl = {https://ui.adsabs.harvard.edu/abs/2018ApJ...869..124S}
}

@ARTICLE{Gonzalez_assembly_2005,
       author = {{Gonzalez}, Anthony H. and {Tran}, Kim-Vy H. and {Conbere}, Michelle N. and {Zaritsky}, Dennis},
        title = "{Galaxy Cluster Assembly at z=0.37}",
      journal = {\apjl},
         year = 2005,
        month = may,
       volume = {624},
       number = {2},
        pages = {L73-L76},
          doi = {10.1086/430518},
archivePrefix = {arXiv},
       eprint = {astro-ph/0503451},
 primaryClass = {astro-ph},
       adsurl = {https://ui.adsabs.harvard.edu/abs/2005ApJ...624L..73G}
}

@ARTICLE{Carrasco_formation_2007,
       author = {{Carrasco}, E.~R. and {Cypriano}, E.~S. and {Neto}, G.~B. Lima and {Cuevas}, H. and {Sodr{\'e}}, Jr., L. and {de Oliveira}, C. Mendes and {Ramirez}, A.},
        title = "{Witnessing the Formation of a Galaxy Cluster at z = 0.485: Optical and X-Ray Properties of RX J1117.4+0743 ([VMF 98] 097)}",
      journal = {\apj},
         year = 2007,
        month = aug,
       volume = {664},
       number = {2},
        pages = {777-790},
          doi = {10.1086/518925},
archivePrefix = {arXiv},
       eprint = {0704.2459},
 primaryClass = {astro-ph},
       adsurl = {https://ui.adsabs.harvard.edu/abs/2007ApJ...664..777C}
}

@ARTICLE{Dawson_MCMAC_2013,
       author = {{Dawson}, William A.},
        title = "{The Dynamics of Merging Clusters: A Monte Carlo Solution Applied to the Bullet and Musket Ball Clusters}",
      journal = {\apj},
         year = 2013,
        month = aug,
       volume = {772},
       number = {2},
          eid = {131},
        pages = {131},
          doi = {10.1088/0004-637X/772/2/131},
archivePrefix = {arXiv},
       eprint = {1210.0014},
 primaryClass = {astro-ph.CO},
       adsurl = {https://ui.adsabs.harvard.edu/abs/2013ApJ...772..131D}
}

@ARTICLE{Dawson_merging_2015,
       author = {{Dawson}, William A. and {Jee}, M. James and {Stroe}, Andra and {Ng}, Y. Karen and {Golovich}, Nathan and {Wittman}, David and {Sobral}, David and {Br{\"u}ggen}, M. and {R{\"o}ttgering}, H.~J.~A. and {van Weeren}, R.~J.},
        title = "{MC$^{2}$: Galaxy Imaging and Redshift Analysis of the Merging Cluster CIZA J2242.8+5301}",
      journal = {\apj},
         year = 2015,
        month = jun,
       volume = {805},
       number = {2},
          eid = {143},
        pages = {143},
          doi = {10.1088/0004-637X/805/2/143},
archivePrefix = {arXiv},
       eprint = {1410.2893},
 primaryClass = {astro-ph.GA},
       adsurl = {https://ui.adsabs.harvard.edu/abs/2015ApJ...805..143D}
}

@ARTICLE{Ferragamo2020,
       author = {{Ferragamo}, A. and {Rubi{\~n}o-Mart{\'\i}n}, J.~A. and {Betancort-Rijo}, J. and {Munari}, E. and {Sartoris}, B. and {Barrena}, R.},
        title = "{Biases in galaxy cluster velocity dispersion and mass estimates in the small N$_{gal}$ regime}",
      journal = {\aap},
         year = 2020,
        month = sep,
       volume = {641},
          eid = {A41},
        pages = {A41},
          doi = {10.1051/0004-6361/201834837},
archivePrefix = {arXiv},
       eprint = {2006.05949},
 primaryClass = {astro-ph.CO},
       adsurl = {https://ui.adsabs.harvard.edu/abs/2020A&A...641A..41F}
}

@ARTICLE{Struble_Binary_1958,
       author = {{Struble}, Mitchell F.},
        title = "{Position-Angle Statistics of the Brightest Binary Galaxies in Abell Clusters}",
      journal = {\aj},
         year = 1988,
        month = nov,
       volume = {96},
        pages = {1534},
          doi = {10.1086/114904},
       adsurl = {https://ui.adsabs.harvard.edu/abs/1988AJ.....96.1534S}
}

@ARTICLE{Struble_Morphological_1987,
       author = {{Struble}, Mitchell F. and {Rood}, Herbert J.},
        title = "{A Catalog of Morphological Properties of the 2712 Abell Clusters}",
      journal = {\apjs},
         year = 1987,
        month = mar,
       volume = {63},
        pages = {555},
          doi = {10.1086/191174},
       adsurl = {https://ui.adsabs.harvard.edu/abs/1987ApJS...63..555S}
}

@ARTICLE{Wen_DESI_2024,
       author = {{Wen}, Z.~L. and {Han}, J.~L.},
        title = "{A Catalog of 1.58 Million Clusters of Galaxies Identified from the DESI Legacy Imaging Surveys}",
      journal = {\apjs},
         year = 2024,
        month = jun,
       volume = {272},
       number = {2},
          eid = {39},
        pages = {39},
          doi = {10.3847/1538-4365/ad409d},
archivePrefix = {arXiv},
       eprint = {2404.02002},
 primaryClass = {astro-ph.CO},
       adsurl = {https://ui.adsabs.harvard.edu/abs/2024ApJS..272...39W}
}

@ARTICLE{Wen_SuperCOSMOS_2018,
       author = {{Wen}, Z.~L. and {Han}, J.~L. and {Yang}, F.},
        title = "{A catalogue of clusters of galaxies identified from all sky surveys of 2MASS, WISE, and SuperCOSMOS}",
      journal = {\mnras},
         year = 2018,
        month = mar,
       volume = {475},
       number = {1},
        pages = {343-352},
          doi = {10.1093/mnras/stx3189},
archivePrefix = {arXiv},
       eprint = {1712.02491},
 primaryClass = {astro-ph.GA},
       adsurl = {https://ui.adsabs.harvard.edu/abs/2018MNRAS.475..343W}
}

@ARTICLE{Wen_unWISE_2022,
       author = {{Wen}, Z.~L. and {Han}, J.~L.},
        title = "{Clusters of galaxies up to z = 1.5 identified from photometric data of the Dark Energy Survey and unWISE}",
      journal = {\mnras},
         year = 2022,
        month = jul,
       volume = {513},
       number = {3},
        pages = {3946-3959},
          doi = {10.1093/mnras/stac1149},
archivePrefix = {arXiv},
       eprint = {2204.11215},
 primaryClass = {astro-ph.CO},
       adsurl = {https://ui.adsabs.harvard.edu/abs/2022MNRAS.513.3946W}
}

@ARTICLE{Zou_DESI_2021,
       author = {{Zou}, Hu and {Gao}, Jinghua and {Xu}, Xin and {Zhou}, Xu and {Ma}, Jun and {Zhou}, Zhimin and {Zhang}, Tianmeng and {Nie}, Jundan and {Wang}, Jiali and {Xue}, Suijian},
        title = "{Galaxy Clusters from the DESI Legacy Imaging Surveys. I. Cluster Detection}",
      journal = {\apjs},
         year = 2021,
        month = apr,
       volume = {253},
       number = {2},
          eid = {56},
        pages = {56},
          doi = {10.3847/1538-4365/abe5b0},
archivePrefix = {arXiv},
       eprint = {2101.12340},
 primaryClass = {astro-ph.GA},
       adsurl = {https://ui.adsabs.harvard.edu/abs/2021ApJS..253...56Z}
}

@ARTICLE{Klein_ACTMCMF_2024,
       author = {{Klein}, M. and {Mohr}, J.~J. and {Davies}, C.~T.},
        title = "{The ACT-DR5 MCMF galaxy cluster catalog}",
      journal = {\aap},
         year = 2024,
        month = oct,
       volume = {690},
          eid = {A322},
        pages = {A322},
          doi = {10.1051/0004-6361/202451203},
archivePrefix = {arXiv},
       eprint = {2406.14754},
 primaryClass = {astro-ph.CO},
       adsurl = {https://ui.adsabs.harvard.edu/abs/2024A&A...690A.322K}
}

@ARTICLE{Bhattacharya_DM_2013,
       author = {{Bhattacharya}, Suman and {Habib}, Salman and {Heitmann}, Katrin and {Vikhlinin}, Alexey},
        title = "{Dark Matter Halo Profiles of Massive Clusters: Theory versus Observations}",
      journal = {\apj},
         year = 2013,
        month = mar,
       volume = {766},
       number = {1},
          eid = {32},
        pages = {32},
          doi = {10.1088/0004-637X/766/1/32},
archivePrefix = {arXiv},
       eprint = {1112.5479},
 primaryClass = {astro-ph.CO},
       adsurl = {https://ui.adsabs.harvard.edu/abs/2013ApJ...766...32B}
}

@ARTICLE{Arnaud_YSZ_2010,
       author = {{Arnaud}, M. and {Pratt}, G.~W. and {Piffaretti}, R. and {B{\"o}hringer}, H. and {Croston}, J.~H. and {Pointecouteau}, E.},
        title = "{The universal galaxy cluster pressure profile from a representative sample of nearby systems (REXCESS) and the Y$_{SZ}$ - M$_{500}$ relation}",
      journal = {\aap},
         year = 2010,
        month = jul,
       volume = {517},
          eid = {A92},
        pages = {A92},
          doi = {10.1051/0004-6361/200913416},
archivePrefix = {arXiv},
       eprint = {0910.1234},
 primaryClass = {astro-ph.CO},
       adsurl = {https://ui.adsabs.harvard.edu/abs/2010A&A...517A..92A}
}

@ARTICLE{hernandez-lang22,
       author = {{Hern{\'a}ndez-Lang}, D. and {Zenteno}, A. and {Diaz-Ocampo}, A. and {Cuevas}, H. and {Clancy}, J. and {Prado}, P.~H. and {Ald{\'a}s}, F. and {Pallero}, D. and {Monteiro-Oliveira}, R. and {G{\'o}mez}, F.~A. and {Ramirez}, Amelia and {Wynter}, J. and {Carrasco}, E.~R. and {Hau}, G.~K.~T. and {Stalder}, B. and {McDonald}, M. and {Bayliss}, M. and {Floyd}, B. and {Garmire}, G. and {Katzenberger}, A. and {Kim}, K.~J. and {Klein}, M. and {Mahler}, G. and {Nilo Castellon}, J.~L. and {Saro}, A. and {Somboonpanyakul}, T.},
        title = "{Clash of Titans: A MUSE dynamical study of the extreme cluster merger SPT-CL J0307-6225}",
      journal = {\mnras},
         year = 2022,
        month = dec,
       volume = {517},
       number = {3},
        pages = {4355-4378},
          doi = {10.1093/mnras/stac2480},
archivePrefix = {arXiv},
       eprint = {2111.15443},
 primaryClass = {astro-ph.GA},
       adsurl = {https://ui.adsabs.harvard.edu/abs/2022MNRAS.517.4355H}
}

@ARTICLE{aldas2025,
       author = {{Ald{\'a}s}, F. and {G{\'o}mez}, Facundo A. and {Vega-Mart{\'\i}nez}, C. and {Zenteno}, A. and {Carrasco}, Eleazar R.},
        title = "{Differences in the physical properties of satellite galaxies within relaxed and disturbed galaxy groups and clusters}",
      journal = {\aap},
         year = 2025,
        month = jul,
       volume = {699},
          eid = {A313},
        pages = {A313},
          doi = {10.1051/0004-6361/202451801},
archivePrefix = {arXiv},
       eprint = {2408.05305},
 primaryClass = {astro-ph.GA},
       adsurl = {https://ui.adsabs.harvard.edu/abs/2025A&A...699A.313A}
}

@ARTICLE{aldas2023,
       author = {{Ald{\'a}s}, Franklin and {Zenteno}, Alfredo and {G{\'o}mez}, Facundo A. and {Hernandez-Lang}, Daniel and {Carrasco}, Eleazar R. and {Vega-Mart{\'\i}nez}, Cristian A. and {Nilo Castell{\'o}n}, J.~L.},
        title = "{Clash of Titans: the impact of cluster mergers in the galaxy cluster red sequence}",
      journal = {\mnras},
         year = 2023,
        month = oct,
       volume = {525},
       number = {2},
        pages = {1769-1778},
          doi = {10.1093/mnras/stad2261},
archivePrefix = {arXiv},
       eprint = {2307.11837},
 primaryClass = {astro-ph.GA},
       adsurl = {https://ui.adsabs.harvard.edu/abs/2023MNRAS.525.1769A}
}

@misc{bulbul24,
      title={The SRG/eROSITA All-Sky Survey: The first catalog of galaxy clusters and groups in the Western Galactic Hemisphere}, 
      author={E. Bulbul and A. Liu and M. Kluge and X. Zhang and J. S. Sanders and Y. E. Bahar and V. Ghirardini and E. Artis and R. Seppi and C. Garrel and M. E. Ramos-Ceja and J. Comparat and F. Balzer and K. Böckmann and M. Brüggen and N. Clerc and K. Dennerl and K. Dolag and M. Freyberg and S. Grandis and D. Gruen and F. Kleinebreil and S. Krippendorf and G. Lamer and A. Merloni and K. Migkas and K. Nandra and F. Pacaud and P. Predehl and T. H. Reiprich and T. Schrabback and A. Veronica and J. Weller and S. Zelmer},
      year={2024},
      eprint={2402.08452},
      archivePrefix={arXiv},
      primaryClass={astro-ph.CO}
}

@ARTICLE{dey19,
       author = {{Dey}, Arjun and {Schlegel}, David J. and {Lang}, Dustin and {Blum}, Robert and {Burleigh}, Kaylan and {Fan}, Xiaohui and {Findlay}, Joseph R. and {Finkbeiner}, Doug and {Herrera}, David and {Juneau}, St{\'e}phanie and {Landriau}, Martin and {Levi}, Michael and {McGreer}, Ian and {Meisner}, Aaron and {Myers}, Adam D. and {Moustakas}, John and {Nugent}, Peter and {Patej}, Anna and {Schlafly}, Edward F. and {Walker}, Alistair R. and {Valdes}, Francisco and {Weaver}, Benjamin A. and {Y{\`e}che}, Christophe and {Zou}, Hu and {Zhou}, Xu and {Abareshi}, Behzad and {Abbott}, T.~M.~C. and {Abolfathi}, Bela and {Aguilera}, C. and {Alam}, Shadab and {Allen}, Lori and {Alvarez}, A. and {Annis}, James and {Ansarinejad}, Behzad and {Aubert}, Marie and {Beechert}, Jacqueline and {Bell}, Eric F. and {BenZvi}, Segev Y. and {Beutler}, Florian and {Bielby}, Richard M. and {Bolton}, Adam S. and {Brice{\~n}o}, C{\'e}sar and {Buckley-Geer}, Elizabeth J. and {Butler}, Karen and {Calamida}, Annalisa and {Carlberg}, Raymond G. and {Carter}, Paul and {Casas}, Ricard and {Castander}, Francisco J. and {Choi}, Yumi and {Comparat}, Johan and {Cukanovaite}, Elena and {Delubac}, Timoth{\'e}e and {DeVries}, Kaitlin and {Dey}, Sharmila and {Dhungana}, Govinda and {Dickinson}, Mark and {Ding}, Zhejie and {Donaldson}, John B. and {Duan}, Yutong and {Duckworth}, Christopher J. and {Eftekharzadeh}, Sarah and {Eisenstein}, Daniel J. and {Etourneau}, Thomas and {Fagrelius}, Parker A. and {Farihi}, Jay and {Fitzpatrick}, Mike and {Font-Ribera}, Andreu and {Fulmer}, Leah and {G{\"a}nsicke}, Boris T. and {Gaztanaga}, Enrique and {George}, Koshy and {Gerdes}, David W. and {Gontcho}, Satya Gontcho A. and {Gorgoni}, Claudio and {Green}, Gregory and {Guy}, Julien and {Harmer}, Diane and {Hernandez}, M. and {Honscheid}, Klaus and {Huang}, Lijuan Wendy and {James}, David J. and {Jannuzi}, Buell T. and {Jiang}, Linhua and {Joyce}, Richard and {Karcher}, Armin and {Karkar}, Sonia and {Kehoe}, Robert and {Kneib}, Jean-Paul and {Kueter-Young}, Andrea and {Lan}, Ting-Wen and {Lauer}, Tod R. and {Le Guillou}, Laurent and {Le Van Suu}, Auguste and {Lee}, Jae Hyeon and {Lesser}, Michael and {Perreault Levasseur}, Laurence and {Li}, Ting S. and {Mann}, Justin L. and {Marshall}, Robert and {Mart{\'\i}nez-V{\'a}zquez}, C.~E. and {Martini}, Paul and {du Mas des Bourboux}, H{\'e}lion and {McManus}, Sean and {Meier}, Tobias Gabriel and {M{\'e}nard}, Brice and {Metcalfe}, Nigel and {Mu{\~n}oz-Guti{\'e}rrez}, Andrea and {Najita}, Joan and {Napier}, Kevin and {Narayan}, Gautham and {Newman}, Jeffrey A. and {Nie}, Jundan and {Nord}, Brian and {Norman}, Dara J. and {Olsen}, Knut A.~G. and {Paat}, Anthony and {Palanque-Delabrouille}, Nathalie and {Peng}, Xiyan and {Poppett}, Claire L. and {Poremba}, Megan R. and {Prakash}, Abhishek and {Rabinowitz}, David and {Raichoor}, Anand and {Rezaie}, Mehdi and {Robertson}, A.~N. and {Roe}, Natalie A. and {Ross}, Ashley J. and {Ross}, Nicholas P. and {Rudnick}, Gregory and {Safonova}, Sasha and {Saha}, Abhijit and {S{\'a}nchez}, F. Javier and {Savary}, Elodie and {Schweiker}, Heidi and {Scott}, Adam and {Seo}, Hee-Jong and {Shan}, Huanyuan and {Silva}, David R. and {Slepian}, Zachary and {Soto}, Christian and {Sprayberry}, David and {Staten}, Ryan and {Stillman}, Coley M. and {Stupak}, Robert J. and {Summers}, David L. and {Sien Tie}, Suk and {Tirado}, H. and {Vargas-Maga{\~n}a}, Mariana and {Vivas}, A. Katherina and {Wechsler}, Risa H. and {Williams}, Doug and {Yang}, Jinyi and {Yang}, Qian and {Yapici}, Tolga and {Zaritsky}, Dennis and {Zenteno}, A. and {Zhang}, Kai and {Zhang}, Tianmeng and {Zhou}, Rongpu and {Zhou}, Zhimin},
        title = "{Overview of the DESI Legacy Imaging Surveys}",
      journal = {\aj},
         year = 2019,
        month = may,
       volume = {157},
       number = {5},
          eid = {168},
        pages = {168},
          doi = {10.3847/1538-3881/ab089d},
archivePrefix = {arXiv},
       eprint = {1804.08657},
 primaryClass = {astro-ph.IM},
       adsurl = {https://ui.adsabs.harvard.edu/abs/2019AJ....157..168D}
}

@ARTICLE{Doubrawa2020,
       author = {{Doubrawa}, L. and {Machado}, R.~E.~G. and {Lagan{\'a}}, T.~F. and
         {Lima Neto}, G.~B. and {Monteiro-Oliveira}, R. and {Cypriano}, E.~S.},
        title = "{Simulations of gas sloshing induced by a newly discovered gas poor substructure in galaxy cluster Abell 1644}",
      journal = {\mnras},
         year = 2020,
        month = may,
       volume = {495},
       number = {2},
        pages = {2022-2034},
          doi = {10.1093/mnras/staa1051},
archivePrefix = {arXiv},
       eprint = {2004.13660},
 primaryClass = {astro-ph.GA},
       adsurl = {https://ui.adsabs.harvard.edu/abs/2020MNRAS.495.2022D}
}

@ARTICLE{drlica-wagner22,
       author = {{Drlica-Wagner}, A. and {Ferguson}, P.~S. and {Adam{\'o}w}, M. and {Aguena}, M. and {Allam}, S. and {Andrade-Oliveira}, F. and {Bacon}, D. and {Bechtol}, K. and {Bell}, E.~F. and {Bertin}, E. and {Bilaji}, P. and {Bocquet}, S. and {Bom}, C.~R. and {Brooks}, D. and {Burke}, D.~L. and {Carballo-Bello}, J.~A. and {Carlin}, J.~L. and {Carnero Rosell}, A. and {Carrasco Kind}, M. and {Carretero}, J. and {Castander}, F.~J. and {Cerny}, W. and {Chang}, C. and {Choi}, Y. and {Conselice}, C. and {Costanzi}, M. and {Crnojevi{\'c}}, D. and {da Costa}, L.~N. and {de Vicente}, J. and {Desai}, S. and {Esteves}, J. and {Everett}, S. and {Ferrero}, I. and {Fitzpatrick}, M. and {Flaugher}, B. and {Friedel}, D. and {Frieman}, J. and {Garc{\'\i}a-Bellido}, J. and {Gatti}, M. and {Gaztanaga}, E. and {Gerdes}, D.~W. and {Gruen}, D. and {Gruendl}, R.~A. and {Gschwend}, J. and {Hartley}, W.~G. and {Hernandez-Lang}, D. and {Hinton}, S.~R. and {Hollowood}, D.~L. and {Honscheid}, K. and {Hughes}, A.~K. and {Jacques}, A. and {James}, D.~J. and {Johnson}, M.~D. and {Kuehn}, K. and {Kuropatkin}, N. and {Lahav}, O. and {Li}, T.~S. and {Lidman}, C. and {Lin}, H. and {March}, M. and {Marshall}, J.~L. and {Mart{\'\i}nez-Delgado}, D. and {Mart{\'\i}nez-V{\'a}zquez}, C.~E. and {Massana}, P. and {Mau}, S. and {McNanna}, M. and {Melchior}, P. and {Menanteau}, F. and {Miller}, A.~E. and {Miquel}, R. and {Mohr}, J.~J. and {Morgan}, R. and {Mutlu-Pakdil}, B. and {Mu{\~n}oz}, R.~R. and {Neilsen}, E.~H. and {Nidever}, D.~L. and {Nikutta}, R. and {Nilo Castellon}, J.~L. and {No{\"e}l}, N.~E.~D. and {Ogando}, R.~L.~C. and {Olsen}, K.~A.~G. and {Pace}, A.~B. and {Palmese}, A. and {Paz-Chinch{\'o}n}, F. and {Pereira}, M.~E.~S. and {Pieres}, A. and {Plazas Malag{\'o}n}, A.~A. and {Prat}, J. and {Riley}, A.~H. and {Rodriguez-Monroy}, M. and {Romer}, A.~K. and {Roodman}, A. and {Sako}, M. and {Sakowska}, J.~D. and {Sanchez}, E. and {S{\'a}nchez}, F.~J. and {Sand}, D.~J. and {Santana-Silva}, L. and {Santiago}, B. and {Schubnell}, M. and {Serrano}, S. and {Sevilla-Noarbe}, I. and {Simon}, J.~D. and {Smith}, M. and {Soares-Santos}, M. and {Stringfellow}, G.~S. and {Suchyta}, E. and {Suson}, D.~J. and {Tan}, C.~Y. and {Tarle}, G. and {Tavangar}, K. and {Thomas}, D. and {To}, C. and {Tollerud}, E.~J. and {Troxel}, M.~A. and {Tucker}, D.~L. and {Varga}, T.~N. and {Vivas}, A.~K. and {Walker}, A.~R. and {Weller}, J. and {Wilkinson}, R.~D. and {Wu}, J.~F. and {Yanny}, B. and {Zaborowski}, E. and {Zenteno}, A. and {Delve Collaboration} and {Des Collaboration} and {Astro Data Lab}},
        title = "{The DECam Local Volume Exploration Survey Data Release 2}",
      journal = {\apjs},
         year = 2022,
        month = aug,
       volume = {261},
       number = {2},
          eid = {38},
        pages = {38},
          doi = {10.3847/1538-4365/ac78eb},
archivePrefix = {arXiv},
       eprint = {2203.16565},
 primaryClass = {astro-ph.IM},
       adsurl = {https://ui.adsabs.harvard.edu/abs/2022ApJS..261...38D}
}

@ARTICLE{lopes18,
       author = {{Lopes}, Paulo A.~A. and {Trevisan}, M. and {Lagan{\'a}}, T.~F. and {Durret}, F. and {Ribeiro}, A.~L.~B. and {Rembold}, S.~B.},
        title = "{Optical substructure and BCG offsets of Sunyaev-Zel'dovich and X-ray-selected galaxy clusters}",
      journal = {\mnras},
         year = 2018,
        month = aug,
       volume = {478},
       number = {4},
        pages = {5473-5490},
          doi = {10.1093/mnras/sty1374},
archivePrefix = {arXiv},
       eprint = {1805.09631},
 primaryClass = {astro-ph.CO},
       adsurl = {https://ui.adsabs.harvard.edu/abs/2018MNRAS.478.5473L}
}

@ARTICLE{Lourenco2023,
       author = {{Louren{\c{c}}o}, Ana C.~C. and {Jaff{\'e}}, Y.~L. and {Vulcani}, B. and {Biviano}, A. and {Poggianti}, B. and {Moretti}, A. and {Kelkar}, K. and {Crossett}, J.~P. and {Gitti}, M. and {Smith}, R. and {Lagan{\'a}}, T.~F. and {Gullieuszik}, M. and {Ignesti}, A. and {McGee}, S. and {Wolter}, A. and {Sonkamble}, S. and {M{\"u}ller}, A.},
        title = "{The effect of cluster dynamical state on ram-pressure stripping}",
      journal = {\mnras},
         year = 2023,
        month = dec,
       volume = {526},
       number = {4},
        pages = {4831-4847},
          doi = {10.1093/mnras/stad2972},
archivePrefix = {arXiv},
       eprint = {2309.15934},
 primaryClass = {astro-ph.GA},
       adsurl = {https://ui.adsabs.harvard.edu/abs/2023MNRAS.526.4831L}
}

@ARTICLE{McPartland2016,
       author = {{McPartland}, Conor and {Ebeling}, Harald and {Roediger}, Elke and {Blumenthal}, Kelly},
        title = "{Jellyfish: the origin and distribution of extreme ram-pressure stripping events in massive galaxy clusters}",
      journal = {\mnras},
         year = 2016,
        month = jan,
       volume = {455},
       number = {3},
        pages = {2994-3008},
          doi = {10.1093/mnras/stv2508},
archivePrefix = {arXiv},
       eprint = {1511.00033},
 primaryClass = {astro-ph.GA},
       adsurl = {https://ui.adsabs.harvard.edu/abs/2016MNRAS.455.2994M}
}

@ARTICLE{Roberts2022,
       author = {{Roberts}, Ian D. and {Lang}, Maojin and {Trotsenko}, Daria and {Bemis}, Ashley R. and {Ellison}, Sara L. and {Lin}, Lihwai and {Pan}, Hsi-An and {Ignesti}, Alessandro and {Leslie}, Sarah and {van Weeren}, Reinout J.},
        title = "{LoTSS Jellyfish Galaxies. IV. Enhanced Star Formation on the Leading Half of Cluster Galaxies and Gas Compression in IC3949}",
      journal = {\apj},
         year = 2022,
        month = dec,
       volume = {941},
       number = {1},
          eid = {77},
        pages = {77},
          doi = {10.3847/1538-4357/ac9e9f},
archivePrefix = {arXiv},
       eprint = {2210.16013},
 primaryClass = {astro-ph.GA},
       adsurl = {https://ui.adsabs.harvard.edu/abs/2022ApJ...941...77R}
}

@ARTICLE{mann12,
       author = {{Mann}, Andrew W. and {Ebeling}, Harald},
        title = "{X-ray-optical classification of cluster mergers and the evolution of the cluster merger fraction}",
      journal = {\mnras},
         year = 2012,
        month = mar,
       volume = {420},
       number = {3},
        pages = {2120-2138},
          doi = {10.1111/j.1365-2966.2011.20170.x},
archivePrefix = {arXiv},
       eprint = {1111.2396},
 primaryClass = {astro-ph.CO},
       adsurl = {https://ui.adsabs.harvard.edu/abs/2012MNRAS.420.2120M}
}

@ARTICLE{Mansheim17,
       author = {{Mansheim}, A.~S. and {Lemaux}, B.~C. and {Tomczak}, A.~R. and {Lubin}, L.~M. and {Rumbaugh}, N. and {Wu}, P. -F. and {Gal}, R.~R. and {Shen}, L. and {Dawson}, W.~A. and {Squires}, G.~K.},
        title = "{Suppressed star formation by a merging cluster system}",
      journal = {\mnras},
         year = 2017,
        month = jul,
       volume = {469},
       number = {1},
        pages = {L20-L25},
          doi = {10.1093/mnrasl/slx041},
archivePrefix = {arXiv},
       eprint = {1705.03468},
 primaryClass = {astro-ph.GA},
       adsurl = {https://ui.adsabs.harvard.edu/abs/2017MNRAS.469L..20M}
}

@ARTICLE{Markevitch07,
   author = {{Markevitch}, M. and {Vikhlinin}, A.},
    title = "{Shocks and cold fronts in galaxy clusters}",
  journal = {\physrep},
   eprint = {astro-ph/0701821},
     year = 2007,
    month = may,
   volume = 443,
    pages = {1-53},
      doi = {10.1016/j.physrep.2007.01.001},
   adsurl = {http://adsabs.harvard.edu/abs/2007PhR...443....1M}
}

@ARTICLE{Molnar16,
       author = {{Molnar}, Sandor},
        title = "{Cluster Physics with Merging Galaxy Clusters}",
      journal = {Frontiers in Astronomy and Space Sciences},
         year = 2016,
        month = feb,
       volume = {2},
          eid = {7},
        pages = {7},
          doi = {10.3389/fspas.2015.00007},
       adsurl = {https://ui.adsabs.harvard.edu/abs/2016FrASS...2....7M}
}

@ARTICLE{Monteiro-Oliveira17b,
   author = {{Monteiro-Oliveira}, R. and {Lima Neto}, G.~B. and {Cypriano}, E.~S. and 
	{Machado}, R.~E.~G. and {Capelato}, H.~V. and {Lagan{\'a}}, T.~F. and 
	{Durret}, F. and {Bagchi}, J.},
    title = "{Weak lensing and spectroscopic analysis of the nearby dissociative merging galaxy cluster Abell 3376}",
  journal = {\mnras},
archivePrefix = "arXiv",
   eprint = {1612.07935},
     year = 2017,
    month = jul,
   volume = 468,
    pages = {4566-4578},
      doi = {10.1093/mnras/stx791},
   adsurl = {http://adsabs.harvard.edu/abs/2017MNRAS.468.4566M}
}

@ARTICLE{Monteiro-Oliveira21,
       author = {{Monteiro-Oliveira}, R. and {Soja}, A.~C. and {Ribeiro}, A.~L.~B. and {Bagchi}, J. and {Sankhyayan}, S. and {Candido}, T.~O. and {Flores}, R.~R.},
        title = "{Probing Saraswati's heart: evaluating the dynamical state of the massive galaxy cluster A2631 through a comprehensive weak-lensing and dynamical analysis}",
      journal = {\mnras},
         year = 2021,
        month = feb,
       volume = {501},
       number = {1},
        pages = {756-768},
          doi = {10.1093/mnras/staa3575},
archivePrefix = {arXiv},
       eprint = {2011.07996},
 primaryClass = {astro-ph.GA},
       adsurl = {https://ui.adsabs.harvard.edu/abs/2021MNRAS.501..756M}
}

@ARTICLE{Monteiro-Oliveira22-Hercules,
       author = {{Monteiro-Oliveira}, R. and {Morell}, D.~F. and {Sampaio}, V.~M. and {Ribeiro}, A.~L.~B. and {de Carvalho}, R.~R.},
        title = "{Unveiling the internal structure of the Hercules supercluster}",
      journal = {\mnras},
         year = 2022,
        month = jan,
       volume = {509},
       number = {3},
        pages = {3470-3487},
          doi = {10.1093/mnras/stab3225},
archivePrefix = {arXiv},
       eprint = {2111.03053},
 primaryClass = {astro-ph.CO},
       adsurl = {https://ui.adsabs.harvard.edu/abs/2022MNRAS.509.3470M}
}

@ARTICLE{Monteiro-Oliveira22-eROSITA,
       author = {{Monteiro-Oliveira}, Rog{\'e}rio},
        title = "{A major galaxy cluster merger caught by eROSITA: weak lensing mass distribution and kinematic description}",
      journal = {\mnras},
         year = 2022,
        month = sep,
       volume = {515},
       number = {3},
        pages = {3674-3684},
          doi = {10.1093/mnras/stac2053},
archivePrefix = {arXiv},
       eprint = {2207.02232},
 primaryClass = {astro-ph.CO},
       adsurl = {https://ui.adsabs.harvard.edu/abs/2022MNRAS.515.3674M}
}

@ARTICLE{nelson24,
       author = {{Nelson}, Dylan and {Pillepich}, Annalisa and {Ayromlou}, Mohammadreza and {Lee}, Wonki and {Lehle}, Katrin and {Rohr}, Eric and {Truong}, Nhut},
        title = "{Introducing the TNG-Cluster simulation: Overview and the physical properties of the gaseous intracluster medium}",
      journal = {\aap},
         year = 2024,
        month = jun,
       volume = {686},
          eid = {A157},
        pages = {A157},
          doi = {10.1051/0004-6361/202348608},
archivePrefix = {arXiv},
       eprint = {2311.06338},
 primaryClass = {astro-ph.GA},
       adsurl = {https://ui.adsabs.harvard.edu/abs/2024A&A...686A.157N}
}

@ARTICLE{Sarazin04,
   author = {{Sarazin}, C.~L.},
    title = "{Mergers, Cosmic Rays, and Nonthermal Processes in Clusters of Galaxies}",
  journal = {Journal of Korean Astronomical Society},
     year = 2004,
    month = dec,
   volume = 37,
    pages = {433-438},
      doi = {10.5303/JKAS.2004.37.5.433},
   adsurl = {http://adsabs.harvard.edu/abs/2004JKAS...37..433S}
}

@ARTICLE{Soja18,
   author = {{Soja}, A.~C. and {Sodr{\'e}}, L. and {Monteiro-Oliveira}, R. and 
	{Cypriano}, E.~S. and {Lima Neto}, G.~B.},
    title = "{A Gemini view of the galaxy cluster RXC J1504-0248: insights on the nature of the central gaseous filaments}",
  journal = {\mnras},
archivePrefix = "arXiv",
   eprint = {1803.03625},
     year = 2018,
    month = jul,
   volume = 477,
    pages = {3279-3292},
      doi = {10.1093/mnras/sty638},
   adsurl = {http://adsabs.harvard.edu/abs/2018MNRAS.477.3279S}
}

@ARTICLE{veliz2025,
       author = {{V{\'e}liz Astudillo}, S. and {Carrasco}, E.~R. and {Nilo Castell{\'o}n}, J.~L. and {Zenteno}, A. and {Cuevas}, H.},
        title = "{The effect of dynamical states on galaxy cluster populations: II. Comparison of galaxy properties and fundamental relations}",
      journal = {\aap},
         year = 2025,
        month = oct,
       volume = {702},
          eid = {A251},
        pages = {A251},
          doi = {10.1051/0004-6361/202555199},
archivePrefix = {arXiv},
       eprint = {2504.13337},
 primaryClass = {astro-ph.GA},
       adsurl = {https://ui.adsabs.harvard.edu/abs/2025A&A...702A.251V}
}

@ARTICLE{wenhan13,
       author = {{Wen}, Z.~L. and {Han}, J.~L.},
        title = "{Substructure and dynamical state of 2092 rich clusters of galaxies derived from photometric data}",
      journal = {\mnras},
         year = 2013,
        month = nov,
       volume = {436},
       number = {1},
        pages = {275-293},
          doi = {10.1093/mnras/stt1581},
archivePrefix = {arXiv},
       eprint = {1307.0568},
 primaryClass = {astro-ph.CO},
       adsurl = {https://ui.adsabs.harvard.edu/abs/2013MNRAS.436..275W}
}

@ARTICLE{Yuan22,
       author = {{Yuan}, Z.~S. and {Han}, J.~L. and {Wen}, Z.~L.},
        title = "{Dynamical state of galaxy clusters evaluated from X-ray images}",
      journal = {\mnras},
         year = 2022,
        month = jun,
       volume = {513},
       number = {2},
        pages = {3013-3021},
          doi = {10.1093/mnras/stac1037},
archivePrefix = {arXiv},
       eprint = {2204.02699},
 primaryClass = {astro-ph.GA},
       adsurl = {https://ui.adsabs.harvard.edu/abs/2022MNRAS.513.3013Y}
}

@ARTICLE{zenteno20,
       author = {{Zenteno}, A. and {Hern{\'a}ndez-Lang}, D. and {Klein}, M. and
         {Vergara Cervantes}, C. and {Hollowood}, D.~L. and {Bhargava}, S. and
         {Palmese}, A. and {Strazzullo}, V. and {Romer}, A.~K. and
         {Mohr}, J.~J. and {Jeltema}, T. and {Saro}, A. and {Lidman}, C. and
         {Gruen}, D. and {Ojeda}, V. and {Katzenberger}, A. and {Aguena}, M. and
         {Allam}, S. and {Avila}, S. and {Bayliss}, M. and {Bertin}, E. and
         {Brooks}, D. and {Buckley-Geer}, E. and {Burke}, D.~L. and
         {Capasso}, R. and {Carnero Rosell}, A. and {Carrasco Kind}, M. and
         {Carretero}, J. and {Castander}, F.~J. and {Costanzi}, M. and
         {da Costa}, L.~N. and {De Vicente}, J. and {Desai}, S. and
         {Diehl}, H.~T. and {Doel}, P. and {Eifler}, T.~F. and {Evrard}, A.~E. and
         {Flaugher}, B. and {Floyd}, B. and {Fosalba}, P. and {Frieman}, J. and
         {Garc{\'\i}a-Bellido}, J. and {Gerdes}, D.~W. and {Gonzalez}, J.~R. and
         {Gruendl}, R.~A. and {Gschwend}, J. and {Gutierrez}, G. and
         {Hartley}, W.~G. and {Hinton}, S.~R. and {Honscheid}, K. and
         {James}, D.~J. and {Kuehn}, K. and {Lahav}, O. and {Lima}, M. and
         {McDonald}, M. and {Maia}, M.~A.~G. and {March}, M. and {Melchior}, P. and
         {Menanteau}, F. and {Miquel}, R. and {Ogando}, R.~L.~C. and
         {Paz-Chinch{\'o}n}, F. and {Plazas}, A.~A. and {Roodman}, A. and
         {Rykoff}, E.~S. and {Sanchez}, E. and {Scarpine}, V. and
         {Schubnell}, M. and {Serrano}, S. and {Sevilla-Noarbe}, I. and
         {Smith}, M. and {Soares-Santos}, M. and {Suchyta}, E. and
         {Swanson}, M.~E.~C. and {Tarle}, G. and {Thomas}, D. and
         {Varga}, T.~N. and {Walker}, A.~R. and {Wilkinson}, R.~D. and
         {DES Collaboration}},
        title = "{A joint SZ-X-ray-optical analysis of the dynamical state of 288 massive galaxy clusters}",
      journal = {\mnras},
         year = 2020,
        month = may,
       volume = {495},
       number = {1},
        pages = {705-725},
          doi = {10.1093/mnras/staa1157},
archivePrefix = {arXiv},
       eprint = {2004.01721},
 primaryClass = {astro-ph.GA},
       adsurl = {https://ui.adsabs.harvard.edu/abs/2020MNRAS.495..705Z}
}

@ARTICLE{Nascimento_Dynamical_2016,
       author = {{Nascimento}, R.~S. and {Ribeiro}, A.~L.~B. and {Trevisan}, M. and {Carrasco}, E.~R. and {Plana}, H. and {Dupke}, R.},
        title = "{Dynamical analysis of the cluster pair: A3407 + A3408}",
      journal = {\mnras},
         year = 2016,
        month = aug,
       volume = {460},
       number = {2},
        pages = {2193-2206},
          doi = {10.1093/mnras/stw1114},
archivePrefix = {arXiv},
       eprint = {1605.02656},
 primaryClass = {astro-ph.GA},
       adsurl = {https://ui.adsabs.harvard.edu/abs/2016MNRAS.460.2193N}
}

@ARTICLE{White_merger_2015,
       author = {{White}, J.~A. and {Canning}, R.~E.~A. and {King}, L.~J. and {Lee}, B.~E. and {Russell}, H.~R. and {Baum}, S.~A. and {Clowe}, D.~I. and {Coleman}, J.~E. and {Donahue}, M. and {Edge}, A.~C. and {Fabian}, A.~C. and {Johnstone}, R.~M. and {McNamara}, B.~R. and {O'Dea}, C.~P. and {Sanders}, J.~S.},
        title = "{Dynamical analysis of galaxy cluster merger Abell 2146}",
      journal = {\mnras},
         year = 2015,
        month = nov,
       volume = {453},
       number = {3},
        pages = {2718-2730},
          doi = {10.1093/mnras/stv1831},
archivePrefix = {arXiv},
       eprint = {1508.01505},
 primaryClass = {astro-ph.CO},
       adsurl = {https://ui.adsabs.harvard.edu/abs/2015MNRAS.453.2718W}
}

@ARTICLE{Araya-Melo_halo_2009,
       author = {{Araya-Melo}, Pablo A. and {van de Weygaert}, Rien and {Jones}, Bernard J.~T.},
        title = "{Cosmology and cluster halo scaling relations}",
      journal = {\mnras},
         year = 2009,
        month = dec,
       volume = {400},
       number = {3},
        pages = {1317-1336},
          doi = {10.1111/j.1365-2966.2009.15565.x},
archivePrefix = {arXiv},
       eprint = {0906.0373},
 primaryClass = {astro-ph.CO},
       adsurl = {https://ui.adsabs.harvard.edu/abs/2009MNRAS.400.1317A}
}

@ARTICLE{Navarro_Universal_1997,
       author = {{Navarro}, Julio F. and {Frenk}, Carlos S. and {White}, Simon D.~M.},
        title = "{A Universal Density Profile from Hierarchical Clustering}",
      journal = {\apj},
         year = 1997,
        month = dec,
       volume = {490},
       number = {2},
        pages = {493-508},
          doi = {10.1086/304888},
archivePrefix = {arXiv},
       eprint = {astro-ph/9611107},
 primaryClass = {astro-ph},
       adsurl = {https://ui.adsabs.harvard.edu/abs/1997ApJ...490..493N}
}

@ARTICLE{Navarro_Halos_1996,
       author = {{Navarro}, Julio F. and {Frenk}, Carlos S. and {White}, Simon D.~M.},
        title = "{The Structure of Cold Dark Matter Halos}",
      journal = {\apj},
         year = 1996,
        month = may,
       volume = {462},
        pages = {563},
          doi = {10.1086/177173},
archivePrefix = {arXiv},
       eprint = {astro-ph/9508025},
 primaryClass = {astro-ph},
       adsurl = {https://ui.adsabs.harvard.edu/abs/1996ApJ...462..563N}
}

@ARTICLE{Hou2009,
       author = {{Hou}, Annie and {Parker}, Laura C. and {Harris}, William E. and {Wilman}, David J.},
        title = "{Statistical Tools for Classifying Galaxy Group Dynamics}",
      journal = {\apj},
         year = 2009,
        month = sep,
       volume = {702},
       number = {2},
        pages = {1199-1210},
          doi = {10.1088/0004-637X/702/2/1199},
archivePrefix = {arXiv},
       eprint = {0908.0938},
 primaryClass = {astro-ph.GA},
       adsurl = {https://ui.adsabs.harvard.edu/abs/2009ApJ...702.1199H}
}

@ARTICLE{Tonry_R_1979,
       author = {{Tonry}, J. and {Davis}, M.},
        title = "{A survey of galaxy redshifts. I. Data reduction techniques.}",
      journal = {\aj},
         year = 1979,
        month = oct,
       volume = {84},
        pages = {1511-1525},
          doi = {10.1086/112569},
       adsurl = {https://ui.adsabs.harvard.edu/abs/1979AJ.....84.1511T}
}

@ARTICLE{Kaldare_RVIDLINES_2003,
       author = {{Kaldare}, Raven and {Colless}, Matthew and {Raychaudhury}, Somak and {Peterson}, B.~A.},
        title = "{FLASH redshift survey - I. Observations and catalogue}",
      journal = {\mnras},
         year = 2003,
        month = mar,
       volume = {339},
       number = {3},
        pages = {652-662},
          doi = {10.1046/j.1365-8711.2003.05695.x},
archivePrefix = {arXiv},
       eprint = {astro-ph/0109415},
 primaryClass = {astro-ph},
       adsurl = {https://ui.adsabs.harvard.edu/abs/2003MNRAS.339..652K}
}

@ARTICLE{Yahil1977,
       author = {{Yahil}, A. and {Vidal}, N.~V.},
        title = "{The Velocity Distribution of Galaxies in Clusters}",
      journal = {\apj},
         year = 1977,
        month = jun,
       volume = {214},
        pages = {347-350},
          doi = {10.1086/155257},
       adsurl = {https://ui.adsabs.harvard.edu/abs/1977ApJ...214..347Y}
}

@ARTICLE{kcorrect,
       author = {{Blanton}, Michael R. and {Roweis}, Sam},
        title = "{K-Corrections and Filter Transformations in the Ultraviolet, Optical, and Near-Infrared}",
      journal = {\aj},
         year = 2007,
        month = feb,
       volume = {133},
       number = {2},
        pages = {734-754},
          doi = {10.1086/510127},
archivePrefix = {arXiv},
       eprint = {astro-ph/0606170},
 primaryClass = {astro-ph},
       adsurl = {https://ui.adsabs.harvard.edu/abs/2007AJ....133..734B}
}

@ARTICLE{ZuHone_MergerCatalog_2018,
       author = {{ZuHone}, J.~A. and {Kowalik}, K. and {{\"O}hman}, E. and {Lau}, E. and {Nagai}, D.},
        title = "{The Galaxy Cluster Merger Catalog: An Online Repository of Mock Observations from Simulated Galaxy Cluster Mergers}",
      journal = {\apjs},
         year = 2018,
        month = jan,
       volume = {234},
       number = {1},
          eid = {4},
        pages = {4},
          doi = {10.3847/1538-4365/aa99db},
archivePrefix = {arXiv},
       eprint = {1609.04121},
 primaryClass = {astro-ph.CO},
       adsurl = {https://ui.adsabs.harvard.edu/abs/2018ApJS..234....4Z}
}

@ARTICLE{Machado_A3376_2013,
       author = {{Machado}, Rubens E.~G. and {Lima Neto}, Gast{\~a}o B.},
        title = "{Simulations of the merging galaxy cluster Abell 3376}",
      journal = {\mnras},
         year = 2013,
        month = apr,
       volume = {430},
       number = {4},
        pages = {3249-3260},
          doi = {10.1093/mnras/stt127},
archivePrefix = {arXiv},
       eprint = {1301.4434},
 primaryClass = {astro-ph.CO},
       adsurl = {https://ui.adsabs.harvard.edu/abs/2013MNRAS.430.3249M}
}

@book{DA86,
	edition = {1},
	title = {Goodness-of-{Fit} {Techniques}},
	isbn = {978-0-203-75306-4},
	url = {https://www.taylorfrancis.com/books/9781351444569},
	doi = {10.1201/9780203753064},
	language = {en},
	urldate = {2026-08-10},
	publisher = {Routledge},
	author = {D'Agostino, RalphB. and Stephens, Michael A.},
	year = {1986},
}

@book{mclust,
	address = {Boca Raton},
	edition = {1},
	title = {Model-{Based} {Clustering}, {Classification}, and {Density} {Estimation} {Using} mclust in {R}},
	isbn = {978-1-003-27796-5},
	url = {https://www.taylorfrancis.com/books/9781003277965},
	doi = {10.1201/9781003277965},
	language = {en},
	urldate = {2026-08-10},
	publisher = {Chapman and Hall/CRC},
	author = {Scrucca, Luca and Fraley, Chris and Murphy, T. Brendan and Adrian E., Raftery},
	month = mar,
	year = {2023},
}

@ARTICLE{Aldas_scaling_2026,
       author = {{Ald{\'a}s}, Franklin and {G{\'o}mez}, Facundo A. and {Springel}, Volker and {Pakmor}, R{\"u}diger and {Saro}, Alex and {Castro}, Tiago},
        title = "{X-ray emission maps and scaling relations in IllustrisTNG and MillenniumTNG: Differences between cluster and group regimes}",
      journal = {arXiv e-prints},
         year = 2026,
        month = jun,
          eid = {arXiv:2606.21495},
        pages = {arXiv:2606.21495},
          doi = {10.48550/arXiv.2606.21495},
archivePrefix = {arXiv},
       eprint = {2606.21495},
 primaryClass = {astro-ph.GA},
       adsurl = {https://ui.adsabs.harvard.edu/abs/2026arXiv260621495A}
}

\end{document}